\documentclass[a4paper,12pt]{article}
\usepackage{jcappub} 
\RequirePackage[top=28mm,bottom=28mm,left=25mm,right=25mm]{geometry}

\usepackage{amssymb}
\usepackage{amsmath}
\usepackage{amsfonts}
\usepackage{amssymb}
\usepackage{float}
\usepackage{soul}
\usepackage[normalem]{ulem}
\usepackage[dvipsnames]{xcolor}

\usepackage[
  sorting=none,
  style=numeric-comp,
  giveninits=true
]{biblatex}
\DeclareUnicodeCharacter{0301}{\'{i}}

\renewbibmacro{in:}{}

\usepackage{academicons}
\usepackage{fontawesome5}
\definecolor{orcidlogocol}{rgb}{0.65, 0.807, 0.223}

\newcommand{\orcid}[1]{$\,$\href{https://orcid.org/#1}{\textcolor{orcidlogocol}{\faOrcid}}}

\def\beq{\begin{equation}}
\def\eeq{\end{equation}}
\def\ber{\begin{eqnarray}}
\def\eer{\end{eqnarray}}
\def\benu{\begin{enumerate}}
\def\eenu{\end{enumerate}}

\def\l{\left}
\def\r{\right}
\def\d{{\rm d}}
\newcommand{\sq}{\lower.25ex\hbox{\large$\Box$}}

\def\f{\frac}

\usepackage{cancel}

\usepackage{framed}

\colorlet{shadecolor}{CadetBlue!5}

\title{Eigenvalue formulation of Stochastic Inflation and application to large perturbation generating inflationary features}

\author[a,b]{Swagat S. Mishra\orcid{0000-0003-4057-145X},}
\author[b]{Edmund J. Copeland\orcid{0000-0003-3959-6051},}
\author[b]{and Anne M. Green\orcid{0000-0002-7135-1671}}

\affiliation[a]{Cosmology, Gravity, and Astroparticle Physics Group, Center for Theoretical Physics of the Universe (CTPU-CGA), Institute for Basic Science, Daejeon, 34126, Korea.}
\affiliation[b]{School of Physics and Astronomy,  University of Nottingham, Nottingham, NG7 2RD, United Kingdom.}

\emailAdd{swagatmishra@ibs.re.kr}
\emailAdd{ed.copeland@nottingham.ac.uk}
\emailAdd{anne.green@nottingham.ac.uk}

\abstract{Stochastic inflation is a powerful technique for calculating the probability distribution function (PDF) of large inflationary perturbations, which may collapse to form Primordial Black Holes. The PDF,  $P({\cal N})$, of the stochastic number of e-folds, ${\cal N}$, satisfies an adjoint Fokker-Planck Equation. We develop a new self-contained  eigenvalue technique which can be used to determine $P({\cal N})$. First we apply this method to the simple case of quantum diffusion along a flat potential without any classical drift.
We recover the expression for the PDF that has previously been found using characteristic functions, with an exponential tail,  and a power-law behaviour, $P({\cal N}) \propto {\cal N}^{-3/2}$, in the intermediate regime between the peak and the tail of the PDF.   
Finally we apply the method to constant drift inflation, in the narrow- and broad-well limits. In the narrow-well limit, there is an analytic solution and the PDF is similar to the drift-free case, with a mildly suppressed tail. In the broad-well limit, determining the full set of eigenvalues and eigenfunctions requires a piecewise construction of the spectrum,  and  the broad-well PDF is qualitatively different, with an enhanced peak and a strongly suppressed tail.}

\begin{document}

\maketitle

\section{Introduction}
\label{sec:Intro}

 Primordial Black Holes (PBHs) may form from the gravitational collapse of large over-densities in the early Universe~\cite{Hawking:1971ei,Carr:1974nx,Carr:1975qj}. A leading dark matter candidate, those in the mass range $10^{17} \, {\rm g} \lesssim M_{\rm PBH} \lesssim 10^{22} \, {\rm g}$ can potentially make up all of the dark matter \cite{Green:2020jor,Carr:2020xqk}. 
 They may also play a role in various other astrophysical and cosmological phenomena \cite{Carr:2023tpt}. For instance, some of the Black Holes (BHs) detected via the emission of gravitational waves when BH binaries merge might be of primordial origin~\cite{Sasaki:2018dmp,LISACosmologyWorkingGroup:2023njw,Yuan:2025avq}. PBHs may serve as the progenitors for the supermassive black holes~\cite{Matteri:2025vnv,Prole:2025snf} observed at high redshifts~\cite{Inayoshi:2025isg}. Light PBHs, with initial mass $M_{\rm PBH} \lesssim 10^{15} \, {\rm g}$, which have Hawking evaporated~\cite{Hawking:1974rv,Hawking:1975vcx,Page:1976df} by the present epoch, may have left various imprints on the early Universe  through their decay products~\cite{Carr:1976zz,MacGibbon:1991tj,Barrow:1990he,Ukwatta:2015iba,Klipfel:2025bvh,Keith:2020jww,Baumann:2007yr,Domenech:2021wkk,}.

  The most popular, and arguably the simplest, PBH formation channel is the collapse of large inflationary density fluctuations shortly after horizon-reentry in the early Universe. 
  PBHs are therefore also an excellent probe of small-scale primordial physics~\cite{Carr:1993aq,Josan:2009qn,Gow:2020bzo}.  Since they form from rare extreme over-densities, their abundance is highly sensitive to the form of the tail of the probability distribution function (PDF) of the perturbations. Therefore, to accurately calculate the PBH mass fraction, it is essential to determine the nature of this tail which is expected to be non-Gaussian ~\cite{Bullock:1996at}. 
  One needs a suitable non-perturbative technique for describing such order unity fluctuations. 

In the single field inflationary paradigm~\cite{Baumann:2009ds}, inflation is sourced by a  (real) scalar field, denoted by $\phi$, with a potential $V(\phi)$. The classical evolution during inflation, as a function of the (classical) number of e-folds of expansion, $N = \int H(t)\,\d t \propto \log a(t)$, with $a(t)$ and $H(t) \equiv \dot{a}(t)/a(t)$ being the scale factor and  Hubble parameter (where $\dot{a} \equiv {\rm d} a/{\rm d} t$),
depends upon the shape of the potential.  Typically, the accelerated expansion of space during inflation is well-approximated by a quasi-de Sitter regime (nearly exponential expansion) during which the inflaton slowly rolls down a potential which is close to flat. Furthermore, for a broad class of inflaton potentials, 
the inflaton velocity remains roughly constant (and small) until the end of inflation~\cite{Martin:2024qnn,Martin:2013tda,Mishra:2024axb}. Such an epoch of slow-terminal motion of the inflaton is governed by what is termed as {\em slow-roll}  inflationary dynamics~\cite{Liddle:1994dx,Baumann:2009ds}.
   
The accelerated expansion of space during inflation amplifies and stretches the 
small-scale quantum fluctuations, both scalar and tensor, to super-Hubble scales, after which  they remain frozen until their subsequent Hubble re-entry post inflation. For 
the class of potentials where the inflaton velocity remains almost constant and small, the resulting power spectra are typically small and nearly scale invariant, thereby providing a causal  mechanism for sourcing the observed primordial (scalar) fluctuations on large, cosmological scales~\cite{Planck:2018nkj,Planck:2018jri,AtacamaCosmologyTelescope:2025blo,SPT-3G:2025bzu}. Hence the statistical correlations of these {\em typical perturbations}  (i.e.~those not in the tails of the PDF) on cosmological scales, are usually computed  within the framework of  linear perturbation theory~\cite{Mukhanov:1990me}, in which the primordial PDF is nearly-Gaussian~\cite{Maldacena:2002vr}.

Stochastic inflation~\cite{Starobinsky:1982ee,Sasaki:1995aw,Lyth:2004gb,Wands:2000dp,Lyth:2005fi} is a powerful framework, particularly for computing the statistical correlations of inflationary fluctuations coarse-grained on length scales much larger than the  Hubble radius \textit{i.e.}, scales with comoving wavenumbers satisfying,  $k \leq \sigma \, aH$, where  the coarse-graining parameter $\sigma$   is a small ($\ll 1$) constant~\cite{Vennin:2015hra}. In this framework, the technique of {\em first passage time analysis}  is used to determine the PDF,  $P({\cal N})$, of the  stochastic number of e-folds ${\cal N}$, which is the number of e-folds of expansion obtained in a given realisation of the stochastic dynamics. 
  The PDF has been shown to satisfy an adjoint Fokker-Planck Equation (FPE)~\cite{Pattison:2017mbe} which can be solved with physically well-motivated boundary conditions.

  The adjoint FPE incorporates contributions to the evolution of the coarse-grained inflaton field (and its conjugate momentum) from  the deterministic dynamics through the {\em classical drift terms}. It also incorporates stochastic diffusion through the {\em quantum noise terms} that are  sourced by the Hubble-exiting small-scale quantum fluctuations of the inflaton (and its conjugate momentum)
~\cite{Starobinsky:1982ee,Vennin:2015hra}.
   The PDF of the coarse-grained curvature perturbation, $P(\zeta_{\rm cg})$,  can then  be determined using the {\em stochastic $\delta{\cal N}$ formalism}~\cite{Fujita:2013cna,Vennin:2015hra,Pattison:2017mbe,Ezquiaga:2019ftu}.  
Stochastic inflation is particularly effective~\cite{Pattison:2017mbe,Ezquiaga:2019ftu,Cespedes:2023aal,Vennin:2024yzl,Ye:2026saa} for computing the non-perturbative tail of the PDF~\cite{Celoria:2021vjw,Cohen:2022clv}, which is important for calculating the abundance of PBHs (and also other rare objects~\cite{Ezquiaga:2022qpw}, such as ultra compact minihalos \cite{Ricotti:2009bs}).

  In  order to generate large perturbations, that form a non-negligible abundance of PBHs, a number of proposals  
  for altering the small-scale inflationary dynamics have been put forward within the single field inflationary paradigm.  These generally involve  
  potentials with PBH-forming 
  features, 
  for instance a near-inflection point like feature~\cite{Garcia-Bellido:2017mdw} in the form of a broad and shallow bump~\cite{Ballesteros:2017fsr,Hertzberg:2017dkh,Bhaumik:2019tvl,Mahbub:2019uhl,Ballesteros:2020qam}, a tiny local feature in the form of a bump~\cite{Atal:2019cdz,Mishra:2019pzq} or a dip in the potential~\cite{Mishra:2019pzq}. Since a large amplification of the primordial perturbations due to the presence of such features typically leads to a significant deviation from the slow-roll dynamics~\cite{Motohashi:2017kbs,Mishra:2019pzq,Karam:2022nym}, it is therefore important to develop the stochastic inflationary formalism beyond slow roll~\cite{Ezquiaga:2018gbw,Ballesteros:2020sre,Pattison:2021oen,Ahmadi:2022lsm,Mishra:2023lhe,Artigas:2025nbm,Briaud:2025ayt}.

 In this paper, as a key step towards accurately calculating the abundance of PBHs, we develop an {\em eigenvalue technique}  (also known as the {\em spectral method}), for determining the behaviour of the large amplitude tail of $P({\cal N})$.  When solving the adjoint FPE, this technique is often combined with other methods, such as the characteristic function approach~\cite{Pattison:2017mbe,Ezquiaga:2019ftu}, in order to determine the full PDF, $P({\cal N})$.  To the best of our knowledge, there is no closed-form eigenvalue solution for the PDF 
in the literature on stochastic inflation. We address this by
developing a concrete and self-contained eigenvalue technique which determines the full PDF (under appropriate boundary conditions), and does not rely on partially incorporating other techniques to obtain the PDF. 

After a brief overview of relevant aspects of the classical and stochastic dynamics of inflation in Sec.~\ref{sec:Inf_Dyn}, in Sec.~\ref{sec:Eigen_tech_formulation} we focus on fully developing the eigenvalue formalism. We then apply this formalism in 
Sec.~\ref{sec:ET_Applications} to calculate the PDF $P({\cal N})$ for two regimes, drift-free diffusion and constant-drift inflation, which are relevant for generating large, PBH forming, perturbations. Finally, we conclude with a discussion in Sec.~\ref{sec:Discussion}. 

Throughout we work in natural units with $c=\hbar =1$ and denote the reduced Planck mass by  $m_p \equiv 1/\sqrt{8\pi G} = 2.43 \times 10^{18}~{\rm GeV}$.

\section{Single-field inflationary dynamics in the presence of a feature}
\label{sec:Inf_Dyn}
In this Section we briefly review the classical and stochastic inflationary dynamics for potentials with a PBH-forming feature, which will be crucial  
for the application of the eigenvalue technique we develop in Sec.~\ref{sec:Eigen_tech_formulation}. To make our analysis as general as possible, we express the dynamics of inflation in terms of the kinematic parameters, $\lbrace H,\,\epsilon_H, \, \eta_H\, \rbrace$, 
defined below in Eqs.~\eqref{eq:epsilon_H} and \eqref{eq:eta_H}, 
rather than the potential  and its derivatives,  $\lbrace V(\phi),\,V_\phi, \, V_{\phi\phi} \rbrace$, where $V_\phi \equiv \d V/\d\phi$ etc..

\subsection{Classical dynamics}
\label{sec:Inf_Dyn_cl}
 The classical dynamics during inflation is conveniently characterised in terms of the two Hubble slow roll parameters $\epsilon_H$ and $\eta_H$ defined as~\cite{Liddle:1994dx}
\ber
\epsilon_H &=& -\f{\d\ln H}{\d N}  = \f{1}{2m_p^2} \l(\f{{\rm d}\phi_{\rm cl}}{{\rm d}N}\r)^2  \,,
\label{eq:epsilon_H}\\
\eta_H &=& -\frac{\ddot{\phi}_{\rm cl}}{H\dot{\phi}_{\rm cl}}  =\epsilon_H  - \frac{1}{2\epsilon_H} \, \frac{{\rm d}\epsilon_H}{{\rm d} N} ~ \l( \, \simeq  - \frac{1}{2\epsilon_H} \, \frac{{\rm d}\epsilon_H}{{\rm d} N} \, \r)  \,,
\label{eq:eta_H}
\eer
where ${\phi}_{\rm cl}(t)$ is the classical value of the inflaton field.
The quasi-de Sitter regime (qdS) of nearly exponential expansion during inflation corresponds to $\epsilon_H \ll 1$. The final approximation for $\eta_H$ in Eq.~(\ref{eq:eta_H}), in parentheses, is valid 
for $\epsilon_H \ll |\eta_H|$, which is typically the case for most PBH forming potentials, as discussed below.  The standard slow-roll conditions,  
\beq
\epsilon_H \ll 1 \, , ~~{\rm and}~~  \l| \eta_H \r| \ll 1 \, ,
\label{eq:SR_conditions}
\eeq 
are satisfied by a wide range of smooth 
potentials~\cite{Martin:2024qnn,Martin:2013tda,Mishra:2024axb}.

To allow a systematic analytical treatment of the system, we assume the background evolution during inflation is qdS like, and described by a series of epochs each with a  nearly constant (but not necessarily the same) $\eta_H$. This {\em constant--$\eta_H$ approach} enabled us to derive analytical expressions for the dynamics of noise terms in stochastic inflation in our  previous work~\cite{Mishra:2023lhe} (a similar analysis was also carried out in an earlier work~\cite{Ballesteros:2020sre}). Furthermore, the models of interest satisfy the hierarchy $\epsilon_H \ll |\eta_H|$ when the inflaton is transiting the PBH-forming feature. During such a constant--$\eta_H$, small $\epsilon_H$ regime, Eq.~(\ref{eq:eta_H}) yields 
\beq
- \frac{1}{2} \, \frac{{\rm d}\ln{\epsilon_H}}{{\rm d} N} \simeq \eta_H \hspace{1.0cm} \Rightarrow \hspace{1.0cm}\epsilon_H(N) = \epsilon_{H,i} \, e^{-2\eta_H N} \, ,
\label{eq:epsilon_H_USR_N}
\eeq
where $\epsilon_{H,i}$ is the value of $\epsilon_H$  at the onset of the epoch ($N=0$). 
In the qdS limit, with $\eta_H$ approximately constant and $\epsilon_H \ll |\eta_H|$, the Hubble parameter can be written, using Eqs.~(\ref{eq:epsilon_H})~and~(\ref{eq:epsilon_H_USR_N}), as
\beq
H(N) = H_i \, \exp{\l[ -\f{\epsilon_{H,i} - \epsilon_H(N)}{2\,\eta_H}\r]} =  H_i \, \exp{\l[- \l( \f{\epsilon_{H,i}}{2\,\eta_H}\r) \l( 1 - e^{-2\eta_H\,N}\r)\r]}
\label{eq:Hubble_qdS_eta}
\eeq
where $H_i$ is the Hubble parameter at $N=0$. Since $\epsilon_{H,i} \ll 1$ in the qdS limit, and the classical evolution across the PBH-forming feature lasts only for a few e-folds, $N < \mathcal{O}\l(10\r)$ with $\eta_H \approx \mathcal{O}(1)$, it follows that $\epsilon_H (N) \ll |\eta_H|$.  In this case, Eq.~(\ref{eq:Hubble_qdS_eta}) implies $H \simeq H_i$. Therefore we will work with the constant-Hubble approximation, which is a standard assumption in the stochastic inflation literature.

Accordingly, the classical evolution of the inflaton ($\phi_{\rm cl}$) and its momentum ($\pi_{\rm cl}$) is described by
\begin{align}
\phi_{\rm cl}(N) &= \phi_{{\rm cl},i} + \f{\pi_{{\rm cl},i}}{\eta_H} \l( 1 - e^{-\eta_H N} \r)\, , \\
 \pi_{\rm cl}(N) &\equiv \f{{\rm d}}{{\rm d}N}\phi_{\rm cl}(N) = \pi_{{\rm cl},i} \, e^{-\eta_H N} \, ,
\label{eq:pi_phi_cl}
\end{align}
where $\phi_{{\rm cl},i}$ and $\pi_{{\rm cl},i}$, are the  values of $\phi_{\rm cl}$ and $\pi_{\rm cl}$ at  $N=0$.
From Eq.~(\ref{eq:pi_phi_cl}), during the constant--$\eta_H$ evolution, the classical  field value and momentum are related by 
\beq
\pi_{\rm cl}(\phi_{\rm cl}) = \pi_{{\rm cl},i} - \eta_H\l(\phi_{\rm cl} - \phi_{{\rm cl},i}  \r) \, ,
\label{eq:CR_phi_pi}
\eeq
which describes the {\em classical phase-space trajectory} during constant--$\eta_H$ inflation, and  plays an important role in the stochastic dynamics of the system.  Throughout we use the terminology `constant--$\eta_H$' to describe a phase where $\eta_H$ has any constant value.  In the early Universe literature, inflation with constant $\eta_H$ is commonly referred to as {\em constant-roll inflation}~\cite{Motohashi:2014ppa,Motohashi:2017aob,Cicciarella:2017nls,Anguelova:2017djf,Morse:2018kda,Motohashi:2019rhu,Motohashi:2025qgd}. Furthermore, in some recent studies of PBH formation from inflation models with a feature in the potential, the term constant-roll inflation has  been used more narrowly to denote the (Wands-dual) phase characterised by a constant $\eta_H < 0$, which occurs after a non-attractor ultra-slow-roll phase with $\eta_H \gtrsim 3$, see Refs.~\cite{Karam:2022nym,Bhatt:2022mmn,Tomberg:2023kli,Ballesteros:2024pwn,Wang:2024xdl}.

\subsection{Quantum diffusion and the stochastic $\delta \mathcal{N}$ formalism}
\label{sec:FPE_ET}
 In stochastic inflation, the long wavelength (super-Hubble) part of the inflaton field and its momentum,  coarse-grained with $k \leq \sigma \, aH$, ($\sigma \ll 1$), are denoted by $\Phi$, and $\Pi$  respectively. The stochastic $\delta{\cal N}$ formalism~\cite{Pattison:2017mbe} relates the coarse grained curvature perturbation,  $\zeta_{\rm cg}$,  to the (stochastic) first-passage number of e-folds, ${\cal N}$, via 
 \beq
\zeta_{\rm cg} \equiv \zeta(\Phi, \, \Pi) = {\cal N} - \langle {\cal N}(\Phi, \, \Pi) \rangle \, ,
\label{eq:zeta_deltaN_stoc}
\eeq
where the stochastic realisation-averaged number of e-folds can be determined from the PDF, $P({\cal N};\Phi,\Pi)$, as  
\beq
\langle {\cal N}(\Phi,\,\Pi) \rangle  = \int_0^{\infty} \,  {\cal N} \, P({\cal N};\Phi,\Pi) \, {\rm d}{\cal N} \, .
\label{eq:N_avg}
\eeq
The PDF, $P({\cal N};\Phi,\Pi)$, satisfies an adjoint Fokker-Planck equation (FPE) of the form~\cite{Pattison:2017mbe,Ezquiaga:2019ftu,Vennin:2020kng}
\beq
 \f{\partial}{\partial {\cal N}} P({\cal N};\Phi,\Pi) =   \l[ D_\phi  \f{\partial}{\partial \Phi} + D_\pi \f{\partial}{\partial \Pi} + \f{1}{2}   \Sigma_{\phi\phi} \f{\partial^2}{\partial \Phi^2} + \Sigma_{\phi\pi}  \f{\partial^2}{\partial \Phi \partial \Pi}  + \f{1}{2}  \Sigma_{\pi\pi} \f{\partial^2}{\partial \Pi^2}\r] P({\cal N};\Phi,\Pi) \, ,
\label{eq:Adj_FPE_Phi_Pi}
\eeq
 where  the classical drift terms $\lbrace D_\phi,\,D_\pi\rbrace$ are defined by~\cite{Vennin:2020kng,Ballesteros:2020sre,Mishra:2023lhe}
\beq
D_{\phi}(\Phi) = \pi_{\rm cl}(\Phi) \, ,  \quad D_{\pi}(\Phi)  = \l[ \epsilon_H(\Phi) - \eta_H(\Phi) \r] \,  \pi_{\rm cl}(\Phi)  \, .
\label{eq:FPE_drift_Phi_Pi}
\eeq
 The contribution to the quantum noise terms from small scale ($k > \sigma\,aH$) fluctuations 
 are characterised by the stochastic noise matrix elements $\Sigma_{\phi\phi},\,\Sigma_{\phi\pi},\,\Sigma_{\pi\pi}$,  which were previously calculated in Refs.~\cite{Ballesteros:2020sre,Mishra:2023lhe}  under the constant--$\eta_H$ approximation. 
 
 For a representative class of PBH-forming features in the literature, we found that shortly after transitioning to  an epoch of constant $\eta_H$, the quantum noise terms of the inflaton and its momentum become strongly correlated and fall exponentially with the number of e-folds~\cite{Ballesteros:2020sre,Ahmadi:2022lsm,Mishra:2023lhe}. Following this the noise terms begin to rise, and eventually become almost constant~\cite{Ballesteros:2020sre,Mishra:2023lhe,Tomberg:2023kli,Jackson:2024aoo}. Therefore, in the subsequent analysis of stochastic dynamics, we will assume the noise terms to be constants, representing their dominant final contribution.

 Since the inflaton and its conjugate momentum noise terms are strongly correlated, the PDF  $P({\cal N};\Phi,\Pi)$,  can be conveniently described by the PDF, $P({\cal N};\Phi)$, of  a single stochastic variable, $\Phi$,  satisfying an adjoint FPE on the classical phase-space trajectory~\cite{Tomberg:2023kli,Noorbala:2024fim}, described by Eq.~(\ref{eq:CR_phi_pi}), given by
\beq
 \f{\partial P({\cal N};\Phi) }{\partial {\cal N}}   =   D_{\phi} \, \f{\partial P({\cal N};\Phi)}{\partial \Phi}  + \f{1}{2} \, \Sigma_{\phi\phi} \, \f{\partial^2 P({\cal N};\Phi)}{\partial \Phi^2}   \equiv {\hat {\cal L}_{\rm FP}^{\dagger}}(\Phi)  P({\cal N};\Phi)    \, ,
\label{eq:Adj_FPE_Phi}
\eeq
where ${\hat {\cal L}_{\rm FP}^{\dagger}}(\Phi)$ is the adjoint Fokker-Planck operator, defined as
 \beq
{\hat {\cal L}_{\rm FP}^{\dagger}}(\Phi) = D_{\phi} \, \f{\partial}{\partial \Phi}  + \f{1}{2} \, \Sigma_{\phi\phi} \, \f{\partial^2}{\partial \Phi^2}  \,,
\label{eq:AdjFP_operator}
 \eeq
while the {\em classical drift} term, as defined in Eq.~(\ref{eq:FPE_drift_Phi_Pi}),  is given by 
\beq
D_{\phi}(\Phi) \equiv \pi_{\rm cl} (\Phi)  = \pm\sqrt{2m_p^2 \, \epsilon_H(\Phi)} \, .
\label{eq:EPE_drift_terms}
\eeq
 The (almost-constant) {\em diffusion coefficient} is the noise-matrix element of the inflaton fluctuations,  which is given by~\cite{Ballesteros:2020sre,Mishra:2023lhe}
\beq
\Sigma_{\phi\phi} =   2^{2 \l( \nu - \f{3}{2}\r)} \, \l[ \f{\Gamma\l(\nu\r)}{\Gamma\l(\f{3}{2}\r)} \r]^2 \, \l(\f{H}{2\pi} \r)^2 \, \sigma^{2\l( -\nu + \f{3}{2}  \r)}\,,
\label{eq:Sig_phiphi_CR}
\eeq
 where $\nu$ during a constant--$\eta_H$ epoch is given by $\nu = \l|3/2 - \eta_H\r|$. 
 
\begin{figure}[!t]
\begin{center}
\includegraphics[width=1\textwidth]{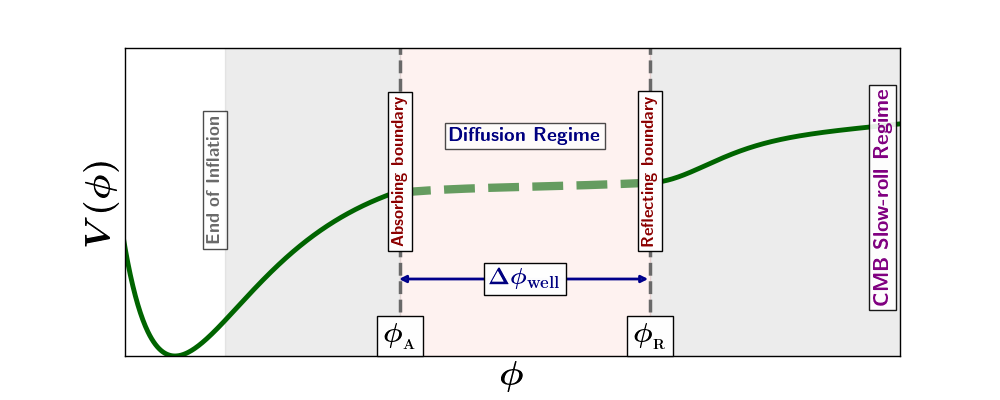}
\caption{A schematic plot in which the green line represents the inflaton potential. The thick dashed-green line denotes an unspecified intermediate feature of width $\Delta\phi_{\rm well} \equiv \phi_{_R} - \phi_{_A}$ in the potential, across which  quantum diffusion becomes important (highlighted with pink shading). After exiting the CMB scale slow-roll phase, the inflaton enters the diffusion regime at the reflecting boundary $\phi = \phi_{_R}$, and later emerges from the diffusion regime at the absorbing boundary $\phi = \phi_{A}$, before the end of inflation.}
\label{fig:inf_SI_feature}
\end{center}
\end{figure}

The adjoint Fokker-Planck Eq.~(\ref{eq:Adj_FPE_Phi}) can be solved using the following two physically motivated boundary conditions~\cite{Pattison:2017mbe}:
 \begin{enumerate}
 \item An absorbing  boundary at $\Phi = \phi_{_A}$,  marking the end of the diffusion regime, where
\beq
P\big\vert_{\Phi \to \phi_{_A}^+}({\cal N};\Phi) = \delta_D({\cal N}) \, ;
\label{eq:BC_IR_PDF}
\eeq
and $\phi_{_A}^+$ indicates that we are approaching $\phi_{_A}$ from field values above it.
 \item A reflecting  boundary at $\Phi = \phi_{_R}$,  preventing the inflaton from diffusing towards arbitrarily large field values (towards the CMB-scale slow-roll regime), where
\beq
\f{\partial}{\partial \Phi}P\big\vert_{ \Phi = \phi_{_R}}({\cal N};\Phi) = 0 \, .
\label{eq:BC_UV_PDF}
\eeq
\end{enumerate}
This set-up is schematically illustrated in Fig.~\ref{fig:inf_SI_feature}, which shows the inflaton potential with an intermediate feature of width $\Delta\phi_{\rm well} \equiv \phi_{_R} - \phi_{_A}$, across which  quantum diffusion of the inflaton becomes important. After exiting the CMB scale slow-roll phase, the inflaton enters the diffusion regime at the reflecting boundary $\phi = \phi_{_R}$, and later emerges from this regime at the absorbing boundary $\phi = \phi_{A}$, before the end of inflation.  

\section{Development  of the eigenvalue technique}
\label{sec:Eigen_tech_formulation}
In the eigenvalue approach, the general solution to the adjoint Fokker-Planck Eq.~(\ref{eq:Adj_FPE_Phi}) is expressed as a spectral decomposition of the form~\cite{Ezquiaga:2019ftu,Markkanen:2019kpv,Ahmadi:2022lsm}
\beq
P({\cal N}; \, \Phi) = \sum_{n} \, c_n \, \Psi_n(\Phi) \, e^{-\Lambda_n \, {\cal N}} \, ,
 \label{eq:PDF_gen_form_Phi}
 \eeq
where $n$ is an integer, and the exponential term, $e^{-\Lambda_n \, {\cal N}}$, is a consequence of the single derivative of the PDF with respect to ${\cal N}$ in Eq.~\eqref{eq:Adj_FPE_Phi}. This is structurally analogous to the single time derivative in the Schr\"odinger equation. The eigenfunctions, $\Psi_n(\Phi)$, satisfy an eigenvalue equation of the form
\beq
{\hat {\cal L}_{\rm FP}^{\dagger}} \Psi_n(\Phi) \equiv   \l[  D_{\phi}  \f{\partial}{\partial \Phi}   +   \f{1}{2}  \Sigma_{\phi\phi}  \f{\partial^2}{\partial \Phi^2}   \r] \Psi_n(\Phi)   = -  \Lambda_n  \Psi_n(\Phi) \,, 
\label{eq:FPE_eigen_gen_Phi}
\eeq 
which, assuming $\Sigma_{\phi\phi} \neq 0$, can also be written as 
\beq
 \f{{\rm d}^2 \Psi_n}{{\rm d} \Phi^2} + \f{2  D_\phi}{\Sigma_{\phi\phi}} \, \f{{\rm d} \Psi_n}{{\rm d} \Phi} + \f{2 \Lambda_n}{\Sigma_{\phi\phi}} \, \Psi_n  = 0   \, ,
\label{eq:FPE_Phi_n}
\eeq 
 with the following two boundary conditions for all $n$\,: an absorbing boundary at $\Phi = \phi_{_A}$ where all the eigenfunctions vanish, namely
\beq
 \Psi_n(\phi_{_A}) = 0\, , 
\label{eq:eigen_BCs_gen_Phi}
\eeq
and a reflecting boundary at $\Phi = \phi_{_R}$ where derivatives of all the eigenfunctions vanish, namely,
\beq
\f{\d}{\d \Phi} \, \Psi_n(\phi_{_R}) = 0\, .
\label{eq:eigen_BCs_gen_Phi_der}
\eeq
The adjoint Fokker-Planck operator ${\hat {\cal L}_{\rm FP}^{\dagger}}(\Phi)$ turns out to be  self-adjoint with respect to a {\em modified inner product} of any two solutions ${\cal F}(\Phi)$ and ${\cal G}(\Phi)$ 
of Eq.~(\ref{eq:FPE_eigen_gen_Phi}), defined as
\beq
\langle {\cal F}, {\cal G}\rangle_w \equiv \int_{\phi_{_A}}^{\phi_{_R}} \, {\rm d}\Phi \, {\cal F}(\Phi) {\cal G}(\Phi) w(\Phi) \, ,
\label{eq:Inner_product}
\eeq
 where $w(\Phi)$ is the {\em Sturm-Liouville weight function}~\cite{SDE_Gardiner} given by (see App.~\ref{app:proof_self_adj_L_FP})
\beq
w(\Phi) =  w_0\,\exp{\l(\f{2 }{\Sigma_{\phi\phi}}\int_{\phi_{_A}}^\Phi \d\Phi \, \, D_\phi(\Phi)\r)}  \,,
\label{eq:w_phi_final}
\eeq 
 where $w_0 = w(\phi_{_A})$ is a constant.

 Consequently, the orthonormality condition for the eigenfunctions becomes
\beq
\langle \Psi_n, \Psi_m \rangle_w \equiv  \int_{\phi_{_A}}^{\phi_{_R}} \, {\rm d}\Phi \, w(\Phi)\,\Psi_n(\Phi) \Psi_m(\Phi) = \delta_{mn} \, .
\label{eq:Psi_n_othonormal_correct}
\eeq
Note that the absorbing boundary condition,  Eq.~(\ref{eq:eigen_BCs_gen_Phi}), determines the functional form of the eigenfunctions $\Psi_n$, while the orthonormality condition, Eq.~(\ref{eq:Psi_n_othonormal_correct}), fixes their normalisation. The reflecting boundary condition, Eq.~(\ref{eq:eigen_BCs_gen_Phi_der}), then  yields the quantized exponents $\Lambda_n$~\cite{Pattison:2017mbe,Ezquiaga:2019ftu}.  Furthermore, the eigenfunctions $\lbrace \Psi_n(\Phi)\rbrace$ form a complete set with respect to the weight function $w(\Phi)$, therefore they satisfy the corresponding completeness relation. Together with the presence of the reflecting boundary at $\phi_{_R}$, this ensures that starting with any $\Phi \in \l[\phi_{_A}, \, \phi_{_R}\r]$, the absorbing boundary at $\phi_{_A}$ is reached with unit probability, so that the first-passage PDF is properly normalized~\cite{Redner,Risken}.

In order to determine the PDF in Eq.~\eqref{eq:PDF_gen_form_Phi}, we need to find the coefficients $c_n$. We do this by specifying the `{\em initial condition}' $P({\cal N}=0; \, \Phi)$, in addition to the boundary conditions given in Eqs.~(\ref{eq:eigen_BCs_gen_Phi})~and~\eqref{eq:eigen_BCs_gen_Phi_der},  and the orthonormality condition,  Eq.~(\ref{eq:Psi_n_othonormal_correct}), in order to formulate a well-defined boundary-value problem~\cite{MR1625845} for the partial differential equation for $P({\cal N}; \, \Phi)$, Eq.~(\ref{eq:Adj_FPE_Phi}). 

As discussed in Sec.~\ref{sec:FPE_ET}, for field values tending towards the absorbing boundary, $\Phi \to \phi_{_A}$,  $P({\cal N}; \, \Phi)$ should be a Dirac delta function in ${\cal N}$, as given in  Eq.~(\ref{eq:BC_IR_PDF}). On the other hand, for ${\cal N}\to 0$ the PDF should be sharply peaked around $\Phi =\phi_{_A}$, in order to ensure that there can be no e-folds of inflation between an arbitrary value of $\Phi > \phi_A$ and $\Phi=\phi_A$ as ${\cal N}\to 0$.
Furthermore, the PDF is zero for $\Phi < \phi_{_A}$, hence it is not symmetric around $\Phi = \phi_{_A}$. A PDF with the following functional form meets these requirements:
\beq
\lim_{{\cal N} \to 0}  P({\cal N}; \, \Phi) \propto  \f{\d}{\d\Phi}\l[  \lim_{{\cal N} \to 0} \,F_s\l(\f{\Phi-\phi_{_A}}{\Delta({\cal N})}\r) \r] \,,
\label{eq:ini_cond_PDF_function}
\eeq
 where $F_s$ is a symmetric function peaked around $\Phi=\phi_{_A}$, 
  with a width $\Delta({\cal N})$ that vanishes as ${\cal N} \to 0$. Since $F_s$ is symmetric, its derivative vanishes at $\Phi=\phi_{_A}$ (before taking the limit $\Delta({\cal N})\to 0$), satisfying the requirement that the PDF vanishes at $\Phi=\phi_{_A}$ for ${\cal N}\neq 0$. Specifically we choose $F_s$ such that, as  $\Delta({\cal N}) \to 0$, it approaches a Dirac delta function. 
  Consequently, the initial condition for the PDF can be prescribed in terms of the derivative of the Dirac delta function:
\beq
 \lim_{{\cal N} \to 0}  P({\cal N}; \, \Phi) = -{\cal B} \, \f{{\rm d}}{{\rm d} \Phi}\delta_D(\Phi-\phi_{_A}) \,.
\label{eq:ini_cond_PDF_delta_der}
\eeq
  Distributions involving derivatives of the Dirac delta function arise in spectral problems with singular localized interactions in quantum mechanics through nontrivial matching conditions at the singular point~\cite{Griffiths1993,Kurasov1996,Albeverio:1988}. Furthermore Eq.~\eqref{eq:ini_cond_PDF_delta_der} is fully consistent with the behaviour of the  PDF obtained using the characteristic function approach in Ref.~\cite{Pattison:2017mbe}, where the same limiting form emerges as a derived result 
  in the drift-free diffusion case, as discussed below in Sec.~\ref{sec:ET_drift_free}.

 The constant ${\cal B}$, found by normalising the absorbing-boundary PDF in Eq.~\eqref{eq:BC_IR_PDF} (see App.~\ref{app:cn_Fourier_trick}), is given by
\beq
{\cal B} = \Biggl[\lim_{\Phi \to \phi_{_A}}    \sum_{m}  \f{\f{\d}{\d\Phi} \l[w(\Phi) \Psi_m(\Phi)\r]\big\vert_{\Phi = \phi_{_A}} \, \times \Psi_m(\Phi)}{\Lambda_m} \Biggr]^{-1} \, .
\label{eq:To_B}
\eeq
The expression for $c_n$,  the coefficient in the PDF $P({\cal N}; \, \Phi)$ in Eq.~(\ref{eq:PDF_gen_form_Phi}), can be calculated using the expression derived in App.~\ref{app:cn_Fourier_trick} 
\beq
c_n =  {\cal B} \times \f{\d}{\d\Phi} \l[w(\Phi) \Psi_n(\Phi)\r]\big\vert_{\Phi = \phi_{_A}} \, ,
\label{eq:c_n_final}
\eeq
and hence the PDF $P({\cal N}; \, \Phi)$ is given by
\beq
 P({\cal N}; \, \Phi) =  {\cal B} \times  \sum_{n}   \f{\d}{\d\Phi} \l[w(\Phi) \Psi_n(\Phi)\r]\big\vert_{\Phi = \phi_{_A}}  \Psi_n(\Phi) \,  e^{-\Lambda_n \, {\cal N}}  \, .
\label{eq:PDF_Eigen_tech_final_w0}
\eeq

The quantity that is more closely related to the  PDF of the coarse-grained curvature perturbations is the PDF for $\delta{\cal N} \equiv {\cal N} - \langle {\cal N} \rangle$. In our spectral formalism,  $\langle {\cal N} \rangle$  can  be calculated using the expression for the PDF in Eq.~(\ref{eq:PDF_Eigen_tech_final_w0}). 
For completeness, the $m^{\rm th}$ moment of the first passage number of e-folds ${\cal N}$ 
is given by~\cite{Pattison:2017mbe}
$$ \langle {\cal N}^m \rangle \equiv \int_0^\infty {\rm d}{\cal N} \, {\cal N}^m \, P({\cal N}; \, \Phi) = \sum_n \, c_n \, \Psi_n(\Phi) \int_0^\infty {\rm d}{\cal N} \, {\cal N}^m e^{-\Lambda_n {\cal N}} \, ,$$
which, upon a change of variable $y = \Lambda_n{\cal N}$, becomes
$$ \langle {\cal N}^m \rangle  = \sum_n \, \f{c_n}{\Lambda_n^{m+1}} \, \Psi_n(\Phi) \int_0^\infty {\rm d}y \, y^m  \, e^{-y} \, ,$$
 leading to
\beq
\langle {\cal N}^m \rangle = \sum_n  \l( \f{m! \, c_n}{\Lambda_n^{m+1}} \r) \, \Psi_n(\Phi) = {\cal B} \times \sum_n   \l( \f{m! }{\Lambda_n^{m+1}} \r) \, \f{\d}{\d\Phi} \l[w(\Phi) \Psi_n(\Phi)\r]\big\vert_{\Phi = \phi_{_A}} \, \Psi_n(\Phi)\, .
\label{eq:m_Moment_N}
\eeq
Hence the average number of first-passage ($m=1$) e-folds is given by
\beq
 \langle {\cal N} \rangle = \sum_n  \l( \f{c_n}{\Lambda_n^2} \r) \, \Psi_n(\Phi) = {\cal B} \times \sum_n   \l( \f{1 }{\Lambda_n^{2}} \r) \, \f{\d}{\d\Phi} \l[w(\Phi) \Psi_n(\Phi)\r]\big\vert_{\Phi = \phi_{_A}}  \, \Psi_n(\Phi) \, .
\label{eq:1st_Moment_N}
\eeq
This can be compared with the classical number of e-folds of expansion between some given field value,  $\phi_{\rm cl} = \phi$, and the absorbing boundary at  $\phi_{\rm cl} = \phi_{_A}$, which is defined as
\beq
N_{\rm cl} \equiv \int_{\phi}^{\phi_{_A}} \, \f{\d\phi_{\rm cl}}{D_\phi(\phi_{\rm cl})} \, .
\label{eq:N_cl_def}
\eeq
 By replacing ${\cal N}$ with $\delta{\cal N} + \langle {\cal N} \rangle$ in Eq.~(\ref{eq:PDF_Eigen_tech_final_w0}), the PDF for $\delta {\cal N}$  in the eigenvalue formalism becomes 
\beq
 P(\delta{\cal N}; \, \Phi) =   {\cal B} \times  \sum_{n}   \f{\d}{\d\Phi} \l[w(\Phi) \Psi_n(\Phi)\r]\big\vert_{\Phi = \phi_{_A}} \, e^{-\Lambda_n \, \langle {\cal N} \rangle } \, \Psi_n(\Phi) \, e^{-\Lambda_n \,  {\cal \delta N}} \, .
 \label{eq:PDF_gen_form_delta_N_1}
 \eeq
 
\section{Application to specific models}
\label{sec:ET_Applications}
We begin in Sec.~\ref{sec:SI_eigen_f}, by discussing relevant physical limits and analytical approximations to the adjoint Fokker-Planck Eq.~(\ref{eq:FPE_Phi_n}). We then  apply the eigenvalue technique developed in Sec.~\ref{sec:Eigen_tech_formulation} to study the stochastic dynamics of two specific models. We first demonstrate the validity and utility of our technique in Sec.~\ref{sec:ET_drift_free}, where we focus on the drift-free diffusion regime. The stochastic dynamics in this case are easier to follow, and the closed form expression for the PDF has already been obtained using other techniques \cite{Pattison:2017mbe,Ezquiaga:2019ftu}. We then move on to study the constant-drift regime in Sec.~\ref{sec:ET_CD}. 

\subsection{Eigenvalue formalism in terms of dimensionless variables}
\label{sec:SI_eigen_f}

We define the dimensionless stochastic field variable $f$ (originally introduced in Ref.~\cite{Pattison:2017mbe})  as  
\beq
f = \frac{\Phi - \phi_{_A}}{\Delta\phi_{\rm well}} \, , 
\label{eq:def_f}
\eeq
 where $\Delta\phi_{\rm well} \equiv \phi_{_R} - \phi_{_A}$ is the width of the potential well, as shown in Fig.~\ref{fig:inf_SI_feature}.
In terms of $f$ the eigenvalue Eq.~(\ref{eq:FPE_Phi_n}) takes the form
\beq
  \f{{\rm d}^2 \Psi_n}{{\rm d} f^2} + \f{2  D_\phi(f) \, \Delta\phi_{\rm well}}{\Sigma_{\phi\phi}} \, \f{{\rm d} \Psi_n}{{\rm d} f} + 2 \Lambda_n \l(\f{\Delta\phi_{\rm well}^2}{\Sigma_{\phi\phi}}\r) \Psi_n  = 0  \,,
\label{eq:FPE_Psi_n_f}
\eeq 
which can be written as
\beq
  \f{{\rm d}^2 \Psi_n}{{\rm d} f^2} - \alpha(f) \, \f{{\rm d} \Psi_n}{{\rm d} f} + \beta_n \, \Psi_n  = 0  \, ,
\label{eq:FPE_Psi_n_f_alpha_beta}
\eeq 
where $\alpha(f)$ and $\beta_n$ are defined as 
\ber 
\alpha(f) &=&  \f{-2\,D_\phi(f)\,\Delta\phi_{\rm well}}{\Sigma_{\phi\phi}} \, , \label{eq:def_alpha_f} \\
\beta_n &=&  \f{2 \Lambda_n\,\Delta\phi_{\rm well}^2}{\Sigma_{\phi\phi}} \, .
\label{eq:def_betan_f}
\eer
Note that $\alpha(f) \geq 0$, independent of the sign  of  the classical drift $D_\phi(f) \equiv \pi_{\rm cl}(f)$\footnote{This is because when $D_\phi(f) <   0$, classically the value of the inflaton field is decreasing so that $\phi_{_R} > \phi_{_A}$ and $\Delta\phi_{\rm well} \equiv \phi_{_R} - \phi_{_A} >  0$, while if $D_\phi(f) > 0$ classically the value of the inflation field is increasing and $\Delta\phi_{\rm well} <0$, so that in both cases  $\alpha(f) \propto - D_\phi(f) \Delta\phi_{\rm well}   \geq 0$.}. Since $\Sigma_{\phi\phi}$, $\Delta\phi_{\rm well}$ and $\Lambda_n$ are all independent of $f$, Eq.~(\ref{eq:FPE_Psi_n_f_alpha_beta}) is the equation of motion of an oscillator with an $f$-dependent friction term, $\alpha(f)$.

Since we are describing stochastic dynamics on the classical phase-space trajectory, under the constant--$\eta_H$ approximation, the drift term $D_\phi(f)$ can be written, by combining Eq.~(\ref{eq:EPE_drift_terms}) with Eq.~(\ref{eq:CR_phi_pi}), as
\beq
D_\phi(f) = D_\phi(f_i) -  \eta_H \l(f -f_i \r)  \Delta\phi_{\rm well} \, ,
\label{eq:Drift_main}
\eeq
where $f_i$ is the initial  value (at $N=0$) of the dimensionless field variable $f$. This initial value can be taken to be at the reflecting boundary, i.e.~at $\Phi_i = \phi_{_R}$, so that $f_i = 1$.  The initial drift $D_\phi(f_i)$ can be expressed in terms of the initial value of the first slow-roll parameter $\epsilon_H$, using Eq.~(\ref{eq:EPE_drift_terms}), as 
\beq
D_{\phi}(f_i) \equiv \pi_{\rm cl,i} = \pm \sqrt{2\epsilon_{H,i}} \, m_p\, .
\label{eq:Drift_initial}
\eeq
 The stochastic evolution is diffusion dominated for $\Sigma_{\phi\phi} \gg D_\phi^2(f)$, while it is drift dominated for $\Sigma_{\phi\phi} \ll D_\phi^2(f)$.

An  important observation from the eigenvalue Eq.~(\ref{eq:FPE_Psi_n_f_alpha_beta}),  
in conjunction with Eqs.~(\ref{eq:def_alpha_f})-(\ref{eq:Drift_initial}), is that the stochastic dynamics is governed by four different (constant) parameters  in the constant--$\eta_H$ approximation, namely, $\lbrace \eta_H, \pi_{\rm cl,i},\,  \Sigma_{\phi\phi}, \,  \Delta\phi_{\rm well} \rbrace$. The parameters $\eta_H$ and $\pi_{\rm cl,i}$ are related to the classical drift, while the noise matrix element $\Sigma_{\phi\phi}$ quantifies stochastic diffusion, and the width $\Delta\phi_{\rm well}$ specifies the field domain associated with the stochastic dynamics. It is possible to construct new parameters by combining these constants in different ways.  In particular, the parameter which will be useful in our analysis  is the {\em dimensionless diffusion width} $\varepsilon$, defined as
\beq
  \varepsilon = \f{\Delta\phi^2_{\rm well}}{\Sigma_{\phi\phi}} \, .
\label{eq:def_varepsilon}
\eeq

As we are working under the constant--$\eta_H$ approximation, Eq.~(\ref{eq:FPE_Psi_n_f_alpha_beta}) can be solved for  two different types of features, namely:
\begin{enumerate}
\item Features with vanishing-$\eta_H$, leading to $D_\phi(f) = D_\phi = \pi_{\rm cl,i}={\rm const}$, which is the case we consider in the present paper. 

In this regime, if 
$\pi_{\rm cl,i} = 0$, which corresponds to $\alpha(f)=0$ in Eq.~(\ref{eq:FPE_Psi_n_f_alpha_beta}),  we refer to it as {\em drift-free diffusion}. This represents the simplest stochastic case, and the key parameter is the dimensionless diffusion width,  $\varepsilon$, defined in Eq.~(\ref{eq:def_varepsilon}).

On the other hand, if $\pi_{\rm cl,i} \neq 0$, which corresponds to $\alpha(f) = {\rm const}$ in Eq.~(\ref{eq:FPE_Psi_n_f_alpha_beta}), we refer to it as {\em constant-drift} inflation, and in this case the key parameter in determining the functional form of the eigenfunctions, and hence of the PDF, is $\alpha$ itself given by Eq.~\eqref{eq:def_alpha_f}. 
One can then take physically-motivated limits of the PDF, such as the drift and diffusion dominated limits, or the broad- and narrow-well approximations.

\item  Features with a constant non-vanishing $\eta_H$ which can lead to a dynamical $\alpha(f)$, with two particularly interesting limits being $\alpha(f) \gg 1$ and $\alpha(f) \ll 1$. Such a constant--$\eta_H$ epoch is an important regime in a number of single field PBH-forming potentials~\cite{Karam:2022nym,Tomberg:2023kli}, as discussed in  Sec.~\ref{sec:Inf_Dyn_cl}. However, the application of the eigenvalue technique to compute the PDF for constant--$\eta_H$ inflation is quite involved, and we defer this to a future publication.
\end{enumerate}

To complete the definitions we require in terms of $f$, we note that by
denoting 
$\Psi_n'(f) \equiv \d\Psi_n/\d f$
the absorbing and reflecting boundary conditions on the eigenfunction, Eq.~\eqref{eq:eigen_BCs_gen_Phi} and Eq.~\eqref{eq:eigen_BCs_gen_Phi_der}, take the form
\ber
 \Psi_n(f)\bigg\vert_{f=0} &=& 0 \, , \label{eq:eigen_BCs_f_Abs} \\
 \Psi_n'(f) \bigg\vert_{f=1} &=& 0  \label{eq:eigen_BCs_f_Ref} \, .
\eer
 Similarly, the orthonormality condition, Eq.~(\ref{eq:Psi_n_othonormal_correct}), for the eigenfunctions becomes
\beq
  \int_{0}^{1} \, {\rm d}f ~ w(f)\,\Psi_n(f) \Psi_m(f) = \f{\delta_{mn}}{\Delta\phi_{\rm well}} \, ,
\label{eq:Psi_n_othonormal_f}
\eeq 
where the weight function, given in Eq.~\eqref{eq:w_phi_final}, can be written as, 
\beq
w(f) =  w_0\,\exp{\l[\f{2 \, \Delta\phi_{\rm well}}{\Sigma_{\phi\phi}}\int_{0}^f \d f\,D_\phi(f)\r]} = w_0\,\exp{\l[-\int_{0}^f \d f\,\alpha(f)\r]}   \,,
\label{eq:w_f_final}
\eeq
which, for the case of constant $D_\phi$ gets reduced to
\beq
w(f) = w_0 \, \exp{\l(\f{2 \, D_\phi\,\Delta\phi_{\rm well}}{\Sigma_{\phi\phi}} \,f\r)} = w_0 \, e^{-\alpha \,f}  \,.
\label{eq:w_f_CD}
\eeq

Consequently, the expression for the PDF of ${\cal N}$ from Eq.~(\ref{eq:PDF_Eigen_tech_final_w0}),  in terms of the dimensionless field variable $f$, becomes
\beq
 P({\cal N}; \, f) =  \l(\f{{\cal B} }{\Delta\phi_{\rm well}}\r) \times  \sum_{n}   \f{\d}{\d f}\l[w(f)\Psi_n(f)\r]\Big\vert_{f=0} \, \Psi_n(f)  e^{-\Lambda_n \, {\cal N}}  \, ,
\label{eq:PDF_N_f}
\eeq
where,  using Eq.~(\ref{eq:To_B}), ${\cal B}$ is given by
\beq
{\cal B} = \Delta\phi_{\rm well} \times  \Biggl[ \lim_{f \to 0}   \sum_{m} \f{ \f{\d}{\d f}\l[w(f)\Psi_m(f)\r]\Big\vert_{f=0} \, \times \Psi_m(f)}{\Lambda_m} \Biggr]^{-1} \, . 
\label{eq:To_B_f}
\eeq
Similarly, 
the expression for $P(\delta{\cal N})$ from Eq.~(\ref{eq:PDF_gen_form_delta_N_1})
becomes 
\beq
 P(\delta{\cal N}; \, f) = \l(\f{{\cal B}}{\Delta\phi_{\rm well}}\r)  \times  \sum_{n}  \f{\d}{\d f}\l[w(f)\Psi_n(f)\r]\Big\vert_{f=0} \, \, e^{-\Lambda_n \, \langle {\cal N} \rangle}  \, \Psi_n(f)  e^{-\Lambda_n \, \delta{\cal N}}  \, ,
 \label{eq:PDF_delta_N_f}
 \eeq
 where $\langle  {\cal N} \rangle$ can be expressed, using Eqs.~(\ref{eq:1st_Moment_N})~and~(\ref{eq:c_n_final}), as
 \beq
\langle {\cal N} \rangle =  \l(\f{ {\cal B} }{\Delta\phi_{\rm well}}\r) \times  \sum_m \f{ \f{\d}{\d f}\l[w(f)\Psi_m(f)\r]\Big\vert_{f=0} \, \Psi_m(f)}{\Lambda_m^2} 
\label{eq:N_avg_f} \,.
 \eeq
In what follows, the PDF is computed numerically by truncating the spectral sum in Eq.~\eqref{eq:PDF_N_f} at a sufficiently large mode number, $n = n_{\rm max}$, with convergence checked over the full parameter range shown in each Figure below. Moreover, given the absorbing boundary condition, Eq.~\eqref{eq:BC_IR_PDF}, the PDF  becomes increasingly sharply peaked at small ${\cal N}$ for $f\to 0$. Consequently, progressively larger values of $n_{\rm max}$ are required to achieve convergence when evaluating the PDF for smaller values of $f$.
\subsection{Drift-free quantum diffusion}
\label{sec:ET_drift_free} 
The simplest case of stochastic dynamics corresponds to quantum diffusion of the inflaton along a flat potential without any classical drift ($D_{\phi}(f)=0$ or equivalently $\alpha(f)=0$).  In this case the eigenvalue Eq.~(\ref{eq:FPE_Psi_n_f})  reduces to that   of a simple harmonic oscillator 
\beq
  \f{{\rm d}^2 \Psi_n}{{\rm d} f^2} + \l( 2 \, \varepsilon \, \Lambda_n \r) \Psi_n  = 0  \,,
\label{eq:FPE_Psi_free_f}
\eeq
where the only free parameter is the dimensionless diffusion width $\varepsilon$, defined in Eq.~(\ref{eq:def_varepsilon}). We solve Eq.~(\ref{eq:FPE_Psi_free_f}) by imposing the absorbing boundary condition, Eq.~(\ref{eq:eigen_BCs_f_Abs}), 
which yields 
\beq
\Psi_n(f) =  A_n \, \sin{ \l( \sqrt{ 2\,\varepsilon\,\Lambda_n} \, f\r) }. 
\label{eq:FPE_free_eigen_f} 
\eeq
Similarly, imposing  the reflecting boundary condition, Eq.~(\ref{eq:eigen_BCs_f_Ref}),
we obtain
$\cos{ \l( \sqrt{ 2\varepsilon\Lambda_n}\r) } = 0$, which yields an expression for the eigenvalues as quantised exponents of the PDF, Eq.~(\ref{eq:PDF_N_f}), 
\beq
\Lambda_n =   \l[ \l( 2n+1 \r)^2 \, \f{\pi^2}{8} \r] \, \f{1}{\varepsilon}  \,,
\label{eq:FPE_free_Lambda}
\eeq
where $n \geq 0$ is a non-negative  integer.  Since $\alpha = 0$ for drift-free diffusion, following Eq.~\eqref{eq:w_f_CD}, the weight function becomes a constant, $w(\Phi) = w_0$. 
Accordingly,  the orthonormality condition, Eq.~(\ref{eq:Psi_n_othonormal_f}), on the eigenfunctions $\Psi_n$ in Eq.~(\ref{eq:FPE_free_eigen_f}),  leads to
\beq
A_n = \sqrt{\f{2}{w_0 \, \Delta\phi_{\rm well}}} \,. 
\label{eq:An_free}
\eeq
 Hence the final exact expression for the  eigenfunction is
\beq
\Psi_n(f) = \sqrt{\f{2}{w_0 \, \Delta\phi_{\rm well}}} \, \sin{ \l[\l( 2n+1\r)\,\f{\pi}{2}\,f \r] }  \, . 
\label{eq:FPE_free_eigen_f_final} 
\eeq
Note that, although $\varepsilon$ was the only free parameter in the eigenvalue Eq.~(\ref{eq:FPE_Psi_free_f}), its solution, Eq.~(\ref{eq:FPE_free_eigen_f_final}),  contains two additional parameters, the width of the well, $\Delta\phi_{\rm well}$, and $w_0$, due to the orthonormality condition, Eq.~(\ref{eq:Psi_n_othonormal_f}). However, as we will see below, the PDF only depends upon the dimensionless width $\varepsilon$.

In order to determine the PDF, Eq.~(\ref{eq:PDF_N_f}), from Eq.~(\ref{eq:FPE_free_eigen_f_final}) we have
\beq
\Psi_n'(f=0) \Psi_n(f) = \l( \f{2}{w_0 \, \Delta\phi_{\rm well}} \r) \,\l[\l( 2n+1\r)\,\f{\pi}{2}\r] \, \sin{ \l[\l( 2n+1\r)\,\f{\pi}{2}\,f \r] }. 
\label{eq:free_psi_der_mult}
\eeq
Consequently, from Eq.~(\ref{eq:To_B_f}), it follows that 
\beq
{\cal B} = \l(\f{\Delta\phi_{\rm well}}{w_0}\r) \, \l(\f{w_0\Delta\phi_{\rm well}}{2} \r) \, \f{1}{\varepsilon} \,  \l[ \lim_{f\to 0} \,  \sum_{m=0}^{\infty}   \,  \f{\sin{ \l[ \l( 2m+1 \r)  \f{\pi}{2} \, f \r] }}{(2m+1)\,\pi/4} \r]^{-1} \, . 
\label{B-drift-free}
\eeq
Using the classic  {\em sine sum Fourier identity}, Eq.~\ref{eq:Sum_sin_odd_1} ~\cite{Gradshteyn:2007},
Eq.~\eqref{B-drift-free} becomes
\beq
\l(\f{w_0}{\Delta\phi_{\rm well}}\r) {\cal B} =  \l(\f{w_0\Delta\phi_{\rm well}}{2} \r) \, \f{1}{\varepsilon} \,.
\label{eq:B_free_final}
\eeq
Hence, using Eqs.~(\ref{eq:free_psi_der_mult})~and~(\ref{eq:B_free_final}), the PDF, Eq.~(\ref{eq:PDF_N_f}), has the final exact form
 \beq
 P({\cal N}; f) =  \l( \f{1}{\varepsilon}\r)  \f{\pi}{2}   \sum_{n=0}^{\infty} \,   (2n+1) \, \sin{ \l[ \l( 2n+1 \r)  \f{\pi}{2} \, f\r] }  \exp{\l[ - \l( 2n+1 \r)^2 \f{\pi^2}{8} \f{1}{\varepsilon} \,  {\cal N}  \r]} \, ,
\label{eq:FPE_free_PDF_final}
\eeq 
which is identical to the results of Refs.~\cite{Pattison:2017mbe,Ezquiaga:2019ftu}, in their case obtained using the {\em characteristic function} approach. The tail of the PDF can be obtained by taking the limit ${\cal N} \, \pi^2/(8\,\varepsilon) \gg 1$. In this case the dominant contribution to Eq.~(\ref{eq:FPE_free_PDF_final}) 
comes from the lowest eigenvalue in Eq.~(\ref{eq:FPE_free_Lambda}), namely,
\beq
\Lambda_0 = \l(\f{\pi^2}{8}\r) \f{1}{\varepsilon} \, ,
\label{eq:PDF_free_tail_exp}
\eeq
which only depends upon $\varepsilon$. Therefore, the tail of the PDF for the drift-free case can be written as 
\beq
P_{\rm Tail}({\cal N}; f) =    \f{\pi}{2\,\varepsilon} \, \sin{ \l(   \f{\pi}{2} \, f\r) } \,   e^{-\f{\pi^2}{8}\l(\f{\cal N}{\varepsilon}\r)} \, .
\label{eq:PDF_free_tail}
\eeq

Note that the PDF in Eq.~\eqref{eq:FPE_free_PDF_final}, expressed as a function of ${\cal N}$ and $f$, depends solely on the dimensionless diffusion width $\varepsilon$ defined in Eq.~\eqref{eq:def_varepsilon}, which is expected since as emphasized earlier, $\varepsilon$ is the only free parameter of the system. Consequently, different choices of the dimensionful physical parameters $\Delta\phi_{\rm well}$ and $\Sigma_{\phi\phi}$ that yield the same ratio $\varepsilon = \Delta\phi_{\rm well}^2/\Sigma_{\phi\phi}$ lead to identical PDFs. 
This behaviour can be understood physically as follows. The diffusion coefficient $\Sigma_{\phi\phi}$, given in Eq.~\eqref{eq:Sig_phiphi_CR}, characterizes the typical amplitude of the quantum jumps experienced by the inflaton during stochastic diffusion~\cite{Starobinsky:1986fx,Vennin:2015hra}, whereas $\Delta\phi_{\rm well}$ is the total width of the quantum well in field space within which the inflaton is allowed to diffuse. Hence, an increase/decrease in $\Sigma_{\phi\phi}$, or equivalently, in the size of the quantum jump, can be compensated by a proportional increase/decrease in the width $\Delta\phi_{\rm well}$, thereby keeping the PDF unchanged.
Therefore, from  Eq.~(\ref{eq:FPE_free_PDF_final}), we can define  a {\em rescaled probability distribution} $\varepsilon \, P({\cal N}; f)$, as a function of the {\em rescaled number of e-folds} ${\cal N}/\varepsilon$, and the dimensionless field  $f$, 
\beq
\varepsilon \, P({\cal N}; f) = \f{\pi}{2}   \sum_{n=0}^{\infty} \,   (2n+1) \, \sin{ \l[ \l( 2n+1 \r)  \f{\pi}{2} \, f\r] }  \exp{\l[ - \l( 2n+1 \r)^2 \f{\pi^2}{8}  \l( \f{ {\cal N}}{\varepsilon} \r) \r]} \, ,
\label{eq:FPE_free_PDF_rescale}
\eeq 
which does not feature any parameter(s).

\begin{figure}[!t]
\begin{center}
\includegraphics[width=0.487\textwidth]{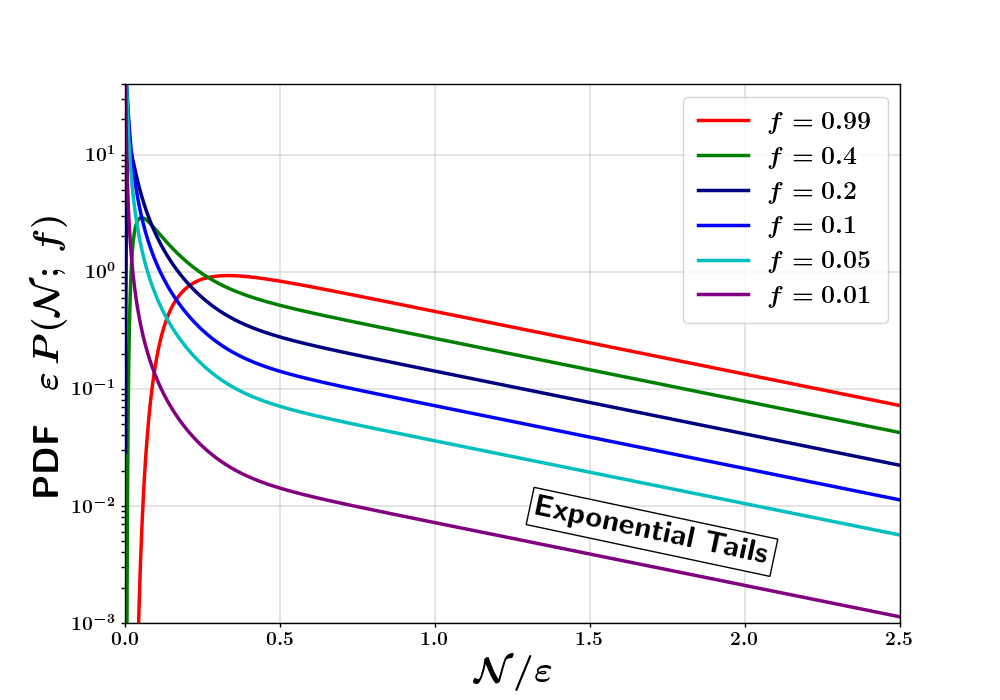}
\includegraphics[width=0.487\textwidth]{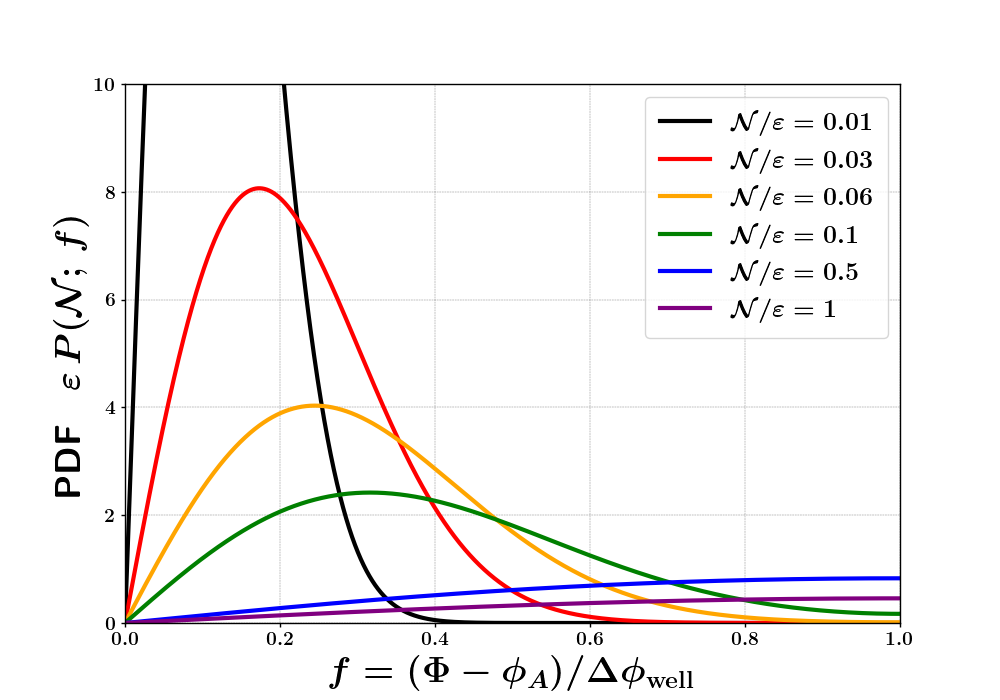}
\caption{The exact rescaled PDF, $\varepsilon \, P({\cal N}; f)$, for the drift-free diffusion case, as given in Eq.~(\ref{eq:FPE_free_PDF_rescale}) as a function of ${\cal N}/ \varepsilon$ for various values of $f$ (left panel), and as a function of $f$  for various values of ${\cal N}/\varepsilon$ (right panel).}
\label{fig:PDF_SR_fQwell}
\end{center}
\end{figure}

The exact rescaled PDF Eq.~(\ref{eq:FPE_free_PDF_rescale}) is plotted in the left panel of Fig.~\ref{fig:PDF_SR_fQwell} as a function of ${\cal N}/\varepsilon$, for  different values of $f$. It is clear that for $f \rightarrow 0$, the rescaled PDF becomes sharply peaked, in accordance with the absorbing boundary condition, Eq.~(\ref{eq:BC_IR_PDF}).
However, for larger values of $f$ the peak in the PDF is smoother. 
Similarly, the right panel shows the PDF as a function of $f$, for  different values of ${\cal N}/\varepsilon$. Note that the PDFs have vanishing slopes at $f=1$, which is a consequence of the reflective  boundary condition, Eq.~(\ref{eq:BC_UV_PDF}).  For the parameter range considered in the case of drift-free diffusion here, we find that summing $\mathcal{O}(10^3)$ modes is sufficient to achieve numerical convergence of the PDF.

\begin{figure}[!t]
\begin{center}
\includegraphics[width=0.487\textwidth]{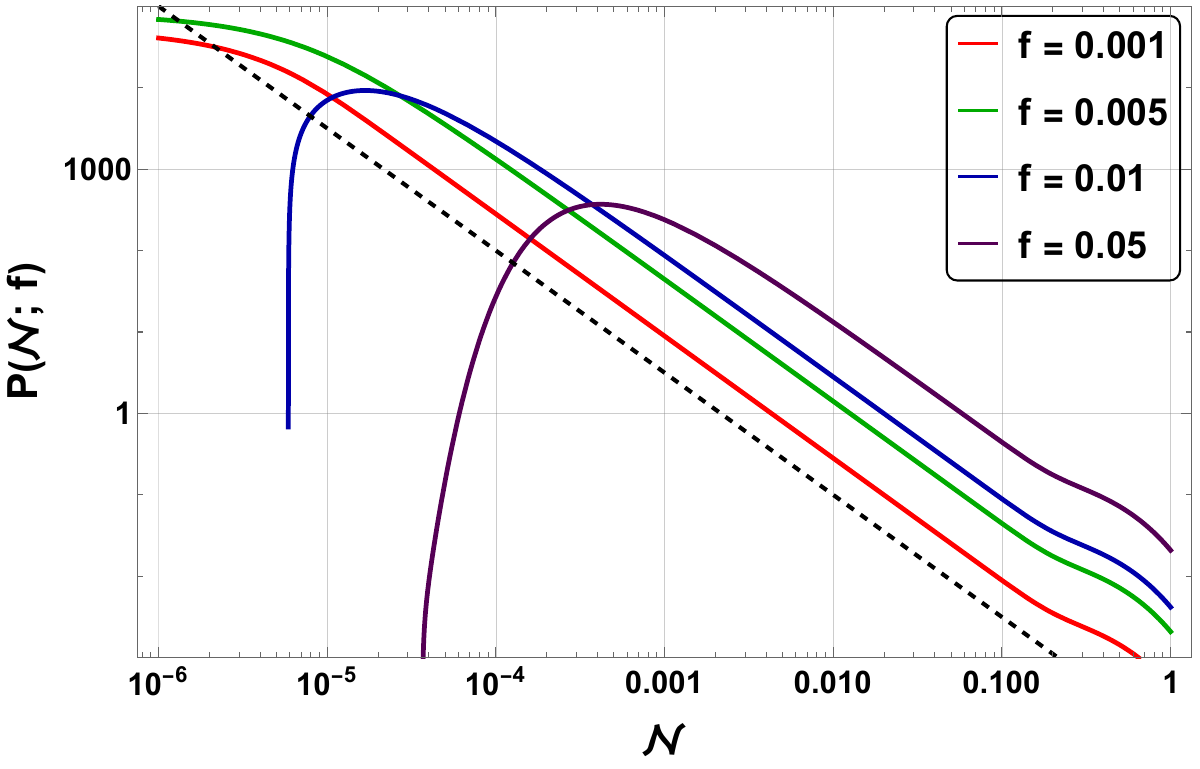}
\includegraphics[width=0.487\textwidth]{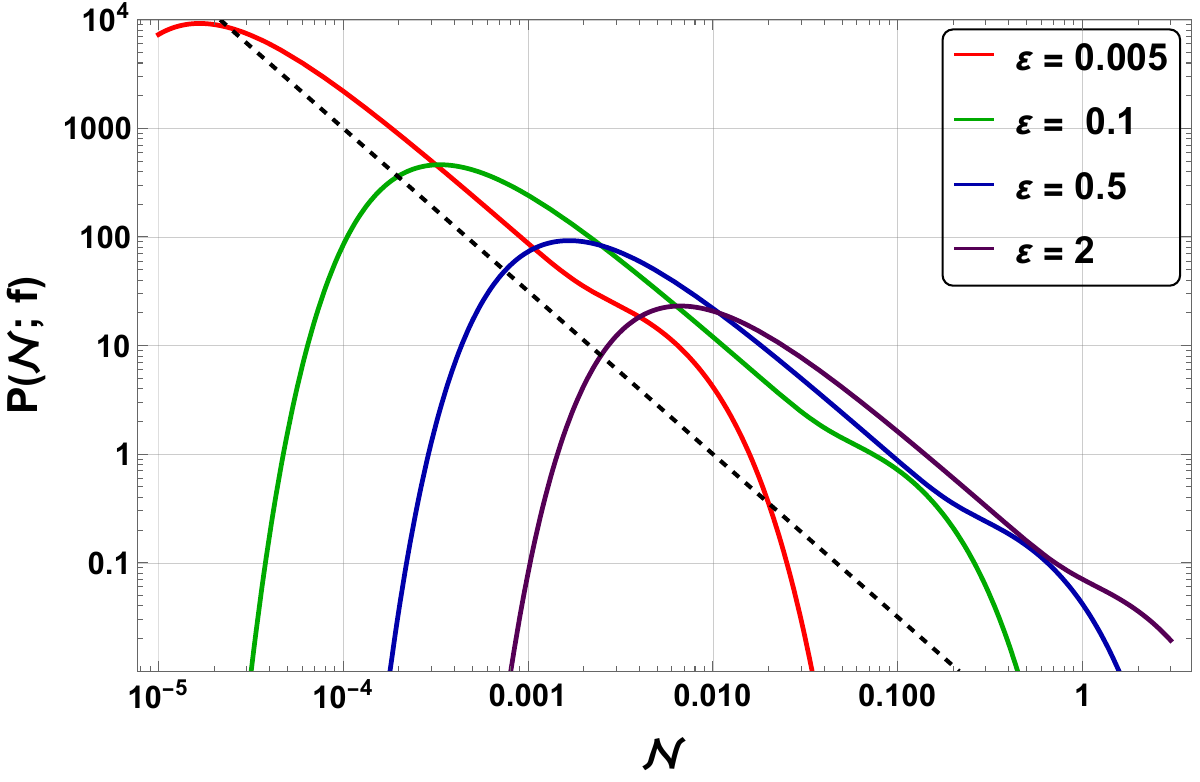}
\caption{The PDF, $P({\cal N}; f)$, for the drift-free diffusion case as a function of ${\cal N}$ for $\varepsilon = 0.5$ and varying $f$ (left panel), and for $f = 0.005$ and varying $\varepsilon$ (right panel). The dashed-black line shows $P({\cal N},\,f) \propto {\cal N}^{-3/2}$.}
\label{fig:PDF_Free_N}
\end{center}
\end{figure}

The PDF in Eq.~(\ref{eq:FPE_free_PDF_final}) can also be expressed  in the following closed form\footnote{This was  originally observed in Ref.~\cite{Pattison:2017mbe}, and was also mentioned in Ref.~\cite{Ezquiaga:2019ftu}.} 
\beq
P({\cal N}; \, f) = - \f{\pi}{4 \, \varepsilon} \, \vartheta_2'\l( \f{\pi}{2} \, f \, ; ~ e^{-\f{\pi^2}{2} \, \f{{\cal N}}{\varepsilon}} \r)\, ,
\label{eq:PDF_free_diff_elliptic}
\eeq
where $\vartheta_2$ is the {\em Jacobi elliptic (theta) function} of the second kind~\cite{book_nist_gov}. Here $\vartheta_2'(X;Y) \equiv \f{\partial}{\partial X}\vartheta_2(X;Y)$. For ${\cal N}/\varepsilon \ll 1$, expanding the elliptic theta function, we obtain the approximate solution 
\beq
P({\cal N}; \, f) \Bigg\vert_{{\cal N}/\varepsilon \ll 1} \approx  \l(\f{\sqrt{\varepsilon}\,f}{\sqrt{2\pi}}\r) \, \f{1}{{\cal N}^{3/2}} \, e^{-\f{\varepsilon f^2}{2{\cal N}}}\, .
\label{eq:PDF_free_diff_elliptic_smallN}
\eeq
 Probability distribution functions of this form are known as {\em Levy distributions}\footnote{We thank Vincent Vennin for pointing out the nomenclature to us.}  in statistics~\cite{albeverio:1988stochastic}.  Note that, as previously found in Ref.~\cite{Pattison:2017mbe}, in the limit ${\cal N}/\varepsilon \to 0^+$, 
\beq
\lim_{{\cal N}/\varepsilon \to 0+} P({\cal N}; \, f) \, \propto \, \f{\d}{\d f} \l[ \lim_{{\cal N}/\varepsilon \to 0+} \l(\f{-1}{\sqrt{2\pi{\cal N}/\varepsilon}}\r) \,  e^{-\f{\varepsilon f^2}{2{\cal N}}} \r] \, = \, -\f{\d}{\d f} \, \delta_D(f)\, ,
\label{eq:Free_der_delta_PDF}
\eeq
which is consistent with our boundary condition in Eq.~(\ref{eq:ini_cond_PDF_delta_der}).  Furthermore, for $f \ll 1$ and $f^2 \ll {\cal N}/\varepsilon$, the above expression reduces to 
\beq
P({\cal N}; \, f)\Bigg\vert_{{\cal N}/\varepsilon \ll 1, f^2 \ll {\cal N}} \approx \l(\f{\sqrt{\varepsilon}\,f}{\sqrt{2\pi}}\r) \, \f{1}{{\cal N}^{3/2}} \, ,
\label{eq:PDF_free_diff_elliptic_smallf}
\eeq
i.e.~$P({\cal N}; \, f) \propto {\cal N}^{-3/2} $.  Such a power-law behaviour of the PDF has also recently been found within the time-reversed stochastic framework~\cite{Blachier:2025tcq} for a tilted semi-infinite potential in Ref.~\cite{Blachier:2025iwk} and, more recently, for a flat quantum well in Ref.~\cite{Animali:2026omc}.
We can see this behaviour in Fig.~\ref{fig:PDF_Free_N} which shows the PDF in Eq.~(\ref{eq:FPE_free_PDF_final}) as a function of ${\cal N}$ for fixed $\varepsilon$ and varying values of $f$ (left panel) and as a function of ${\cal N}$ for fixed $f$ and varying values of $\varepsilon$ (right panel). We see that the PDF interpolates between the peak  and the exponential tail through an intermediate power-law region with $P({\cal N}) \propto {\cal N}^{-3/2}$.
We also find that, while the exponent of the exponential tail depends on the value of $\varepsilon$, the ${\cal N}^{-3/2}$ behaviour of the PDF in this intermediate regime is independent of $\varepsilon$, and hence does not depend on the model parameters $\Sigma_{\phi\phi}$ and $\Delta\phi_{\rm well}$. This intermediate power law tail may affect the abundance of ultra-compact mini halos, that form from slightly smaller perturbations than PBHs.

In order to calculate the PDF of $\delta{\cal N}$, we need to compute the average number of first-passage e-folds $\langle {\cal N}\rangle$, defined in Eq.~\eqref{eq:N_avg_f}. Inserting Eqs.~(\ref{eq:FPE_free_Lambda}), (\ref{eq:free_psi_der_mult}) and (\ref{eq:B_free_final}) into 
Eq.~\eqref{eq:N_avg_f}, we find
\beq 
 \langle {\cal N} \rangle =   \f{32 \, \varepsilon}{\pi^3} \, \sum_{n=0}^{\infty} \,  \f{ 1}{\l( 2n+1 \r)^3}\, \sin{ \l[ \l( 2n+1 \r)  \f{\pi}{2} \, f  \r] }\, .
\label{eq:N_moment1_Free}
\eeq
Using the {\em sine sum identity}, Eq.~(\ref{eq:Sum_sin_odd_3A}) from App.~\ref{app:sums}, 
Eq.~\eqref{eq:N_moment1_Free} reduces to the simple form,
\beq 
\langle {\cal N} \rangle =   2 \, \varepsilon \, f \, \l( 1- \f{f}{2} \r) \, ,
\label{eq:N_moment1_Free1}
\eeq
with the interesting property that $\langle{\cal N}\rangle \propto \varepsilon$. The PDF, $P\l(\delta{\cal N};\,f\r)$, as defined in Eq.~\eqref{eq:PDF_delta_N_f}, can be obtained by replacing  ${\cal N}$ in Eq.~\eqref{eq:FPE_free_PDF_rescale} with $\delta{\cal N} + \langle {\cal N} \rangle$.

    \subsection{Constant-drift inflation}
    \label{sec:ET_CD} 
For stochastic dynamics with a constant drift, \textit{i.e.}~$D_\phi = {\rm constant} ~ (\leq 0)$, we see from Eq.~(\ref{eq:def_alpha_f}) that $\alpha (f) \equiv \alpha = {\rm const}$, hence the eigenvalue Eq.~(\ref{eq:FPE_Psi_n_f_alpha_beta}) takes the form 
\beq
 \f{{\rm d}^2 \Psi_n}{{\rm d} f^2} - \alpha  \,   \f{{\rm d} \Psi_n}{{\rm d} f} + \beta_n \, \Psi_n  = 0  \, ,
\label{eq:FPE_eigen_CD_f}
\eeq 
where the dimensionless parameter 
$\beta_n \equiv 2 \varepsilon \Lambda_n$, is defined in 
Eq.~\eqref{eq:def_betan_f} and is a  constant.  
The general solution to
this equation depends on the sign of the discriminant ${\cal Z}_n^2$ of Eq.~\eqref{eq:FPE_eigen_CD_f}, where 
\beq
 {\cal Z}_n^2 \equiv \beta_n - \f{\alpha^2}{4} \, .
\label{eq:CD_Zn_def}
\eeq 
Note that ${\cal Z}_n^2$ is directly related to the eigenvalue $\Lambda_n$. 
In order to study the stochastic dynamics of the system subject to the boundary conditions, given in Eqs.~\eqref{eq:eigen_BCs_f_Abs}~and~\eqref{eq:eigen_BCs_f_Ref}, we work in the regime ${\cal Z}_n^2 > 0$, since (as we show in App.~\ref{app:CD_Drift-dom}) the case ${\cal Z}_n^2 < 0$ does not result in eigenfunctions that satisfy the boundary conditions.  The condition ${\cal Z}_n^2 > 0$ corresponds to
\beq
{\cal Z}_n^2 > 0 ~ \Leftrightarrow ~ \beta_n > \f{\alpha^2}{4} \,.
\label{eq:CD_Zn_sqr_positive}
\eeq
In this case, the solution to the eigenvalue equation, Eq.~(\ref{eq:FPE_eigen_CD_f}), after imposing the absorbing boundary condition, Eq.~(\ref{eq:eigen_BCs_f_Abs}), at $f=0$, is
\beq
 \Psi_n(f) = A_n \, e^{\f{\alpha}{2} \, f} \,   \sin{ \l({\cal Z}_n f \r) }    \, , \label{eq:FPE_eigen_CD_Abs_sol} 
\eeq
for constant $A_n$.
 Similarly,  imposing the reflecting boundary condition, Eq.~(\ref{eq:eigen_BCs_f_Ref}), at $f = 1$, we obtain a transcendental equation  of the form
\beq
 \tan{ \l( {\cal Z}_n \r) } = -\f{2}{\alpha}  \, {\cal Z}_n \, ,
\label{eq:FPE_CD_Zn_quant}
\eeq
which needs to be solved  for $ {\cal Z}_n$, in order to determine the quantised exponents $\Lambda_n$ using Eq.~(\ref{eq:CD_Zn_def}).  
Given this and using the  orthonormality condition of the eigenfunctions in Eq.~\eqref{eq:Psi_n_othonormal_f}, with the weight function given in Eq.~\eqref{eq:w_f_CD},  we find the exact expressions 
\beq
{\cal B} = \l(\f{ \Delta\phi_{\rm well}}{ 8 \varepsilon w_0}\r) \times  \Biggl[ \lim_{f \to 0}    \sum_{m}  \f{{\cal Z}_m A_m^2 \sin[{\cal Z}_m f]}{\alpha^2 + 4 {\cal Z}_m^2} \Biggr]^{-1} \, ,\label{eq:To_B_f1}
\eeq
where (see App.~\ref{app:CD_calculation_NW} for the derivation) 
\beq
A_n =  \l(\f{2}{w_0\Delta\phi_{\rm well}}\r)^{1/2} \, \, \l( \f{\alpha^2 + 2 \alpha + 4 {\cal Z}_n^2}{\alpha^2  + 4 {\cal Z}_n^2}\r)^{-1/2}\,. 
\label{An1}
\eeq
Therefore, the exact PDF in Eq.~\eqref{eq:PDF_N_f} can be written as
\ber
 P({\cal N}; \, f) = \l(\f{w_0 \, {\cal B}}{\Delta\phi_{\rm well}}\r) \, e^{\f{\alpha}{2}\,f} \,  
 \sum_{n}  A_n^2  \, {\cal Z}_n \,  \sin{\l({\cal Z}_n \, f\r)} \,  e^{-\f{1}{2\varepsilon} \l({\cal Z}_n^2 + \f{\alpha^2}{4} \r) \, {\cal N}}   \, .
\label{eq:PDF_N_f_CD_Gen}
\eer
It proves useful to introduce $\xi_n$ via
\beq
{\cal Z}_n = (n+1) \, \pi - \xi_n  \, , \quad \text{with} \quad 0 \leq \xi_n \leq \f{\pi}{2} \, ,
\label{eq:CD_diff_dom_Zn_sol_new}
\eeq
 where $n \geq 0$ is a non-negative  integer. The limit $\xi_n \to \f{\pi}{2}$ corresponds to the {\em narrow-well regime} ($\alpha  \ll   1$,   $D_\phi \Delta\phi_{\rm well} \ll \Sigma_{\phi\phi}$) and the opposite limit, $\xi_n \to 0$, corresponds to the {\em broad-well regime} ($\alpha  \gg  1 $,  $D_\phi \Delta\phi_{\rm well} \gg \Sigma_{\phi\phi}$).
The exact quantized exponents, $\Lambda_n$, can be written, using Eqs.~(\ref{eq:CD_Zn_def}),~and~(\ref{eq:CD_diff_dom_Zn_sol_new}), as
\beq
 \Lambda_n = \f{1}{2\,\varepsilon}  \left\{ \left[ (n+1) \, \pi - \xi_n \right]^2 +  \l(\f{\alpha}{2}\r)^2 \right\}  \, .
\label{eq:CD_diff_dom_Lambda_n_gen}
\eeq
 Inserting Eq.~(\ref{eq:CD_diff_dom_Zn_sol_new}) into the quantisation condition, Eq.~(\ref{eq:FPE_CD_Zn_quant}), we obtain
\beq
\tan{(\xi_n)} =  \l( \f{2}{\alpha } \r) \Bigl[ (n+1) \, \pi - \xi_n  \Bigr]  \, .
\label{eq:CD_diff_dom_tan_cond_gen}
\eeq
 
\begin{figure}[!t]
\begin{center}
\includegraphics[width=0.75\textwidth]{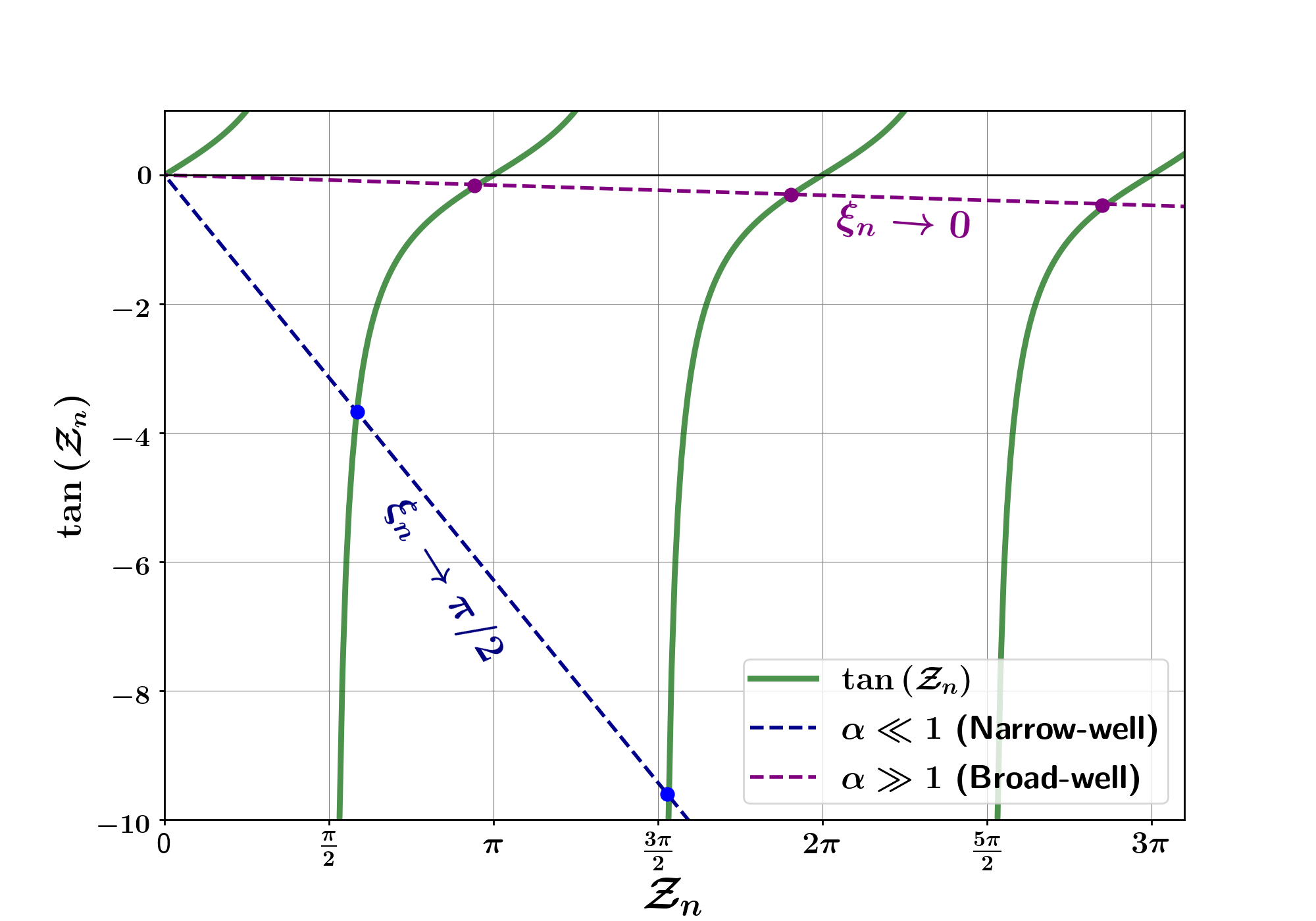}
\caption{An illustration of the solutions of the transcendental equation, Eq.~(\ref{eq:FPE_CD_Zn_quant}), for ${\cal Z}_{n}$ defined in Eq.~(\ref{eq:CD_Zn_def}). The solid green line is $\tan{{\cal Z}_n}$, and the blue and purple circles show the solutions in the narrow ($\alpha \ll 1$) and broad ($\alpha \gg 1$) well limits respectively.}
\label{fig:CD_transc}
\end{center}
\end{figure}

Equation \eqref{eq:CD_diff_dom_tan_cond_gen} is transcendental; it does not admit a closed form analytical solution in terms of elementary functions. 
However, it is instructive to determine $\xi_n$ in  the narrow- and broad-well limits.
The narrow-well limit corresponds to a small (anti-)damping term, $\alpha$, in the eigenvalue Eq.~\eqref{eq:FPE_eigen_CD_f},  which, in the limit $\alpha \to 0$,  reduces to the case of drift-free diffusion discussed in~Sec.~\ref{sec:ET_drift_free}. However, the broad-well limit $\alpha \gg 1$ does not have a natural connection with the drift-free diffusion case. Therefore, quantum diffusion in the broad-well limit of constant-drift inflation offers a distinct, nontrivial system to which we can apply our eigenvalue techniques discussed in Sec.~\ref{sec:Eigen_tech_formulation}.
Fig.~\ref{fig:CD_transc}
shows the solutions to Eq.~(\ref{eq:FPE_CD_Zn_quant}) in these limits. 
Note that the equation for the eigenvalues $\Lambda_n$ written in terms of $\xi_n$, Eq.~\eqref{eq:CD_diff_dom_Lambda_n_gen} 
implies $\Lambda_n > 
 \alpha^2/(8\varepsilon)$, which, as it needs to be, is consistent with Eq.~\eqref{eq:CD_Zn_sqr_positive}, independent of the value of $\xi_n$, and therefore, irrespective of the broad-well or the narrow-well regime \footnote{We note that the asymptotic expansions do not manifestly preserve the normalization of the PDF, however this does not affect its qualitative features.}.

 For constant-drift inflation the classical number of e-folds, Eq.~\eqref{eq:N_cl_def}, which we will  compare to the stochastic average number of e-folds, can be written as
\beq
N_{\rm cl}  = -\f{\Delta\phi_{\rm well}}{D_\phi}\, f =  \l(\f{2\,\varepsilon}{\alpha}\r) f \, .
\label{eq:N_cl_CD}
\eeq
We now apply the eigenvalue techniques to determine the PDF of constant-drift inflation in the narrow and broad-well regimes, in Secs.~\ref{sec:ET_CD_NW} and \ref{sec:ET_CD_BW} respectively.

\subsubsection{Narrow-well limit}
\label{sec:ET_CD_NW}
\begin{figure}[!t]
\begin{center}
\includegraphics[width=0.487\textwidth]{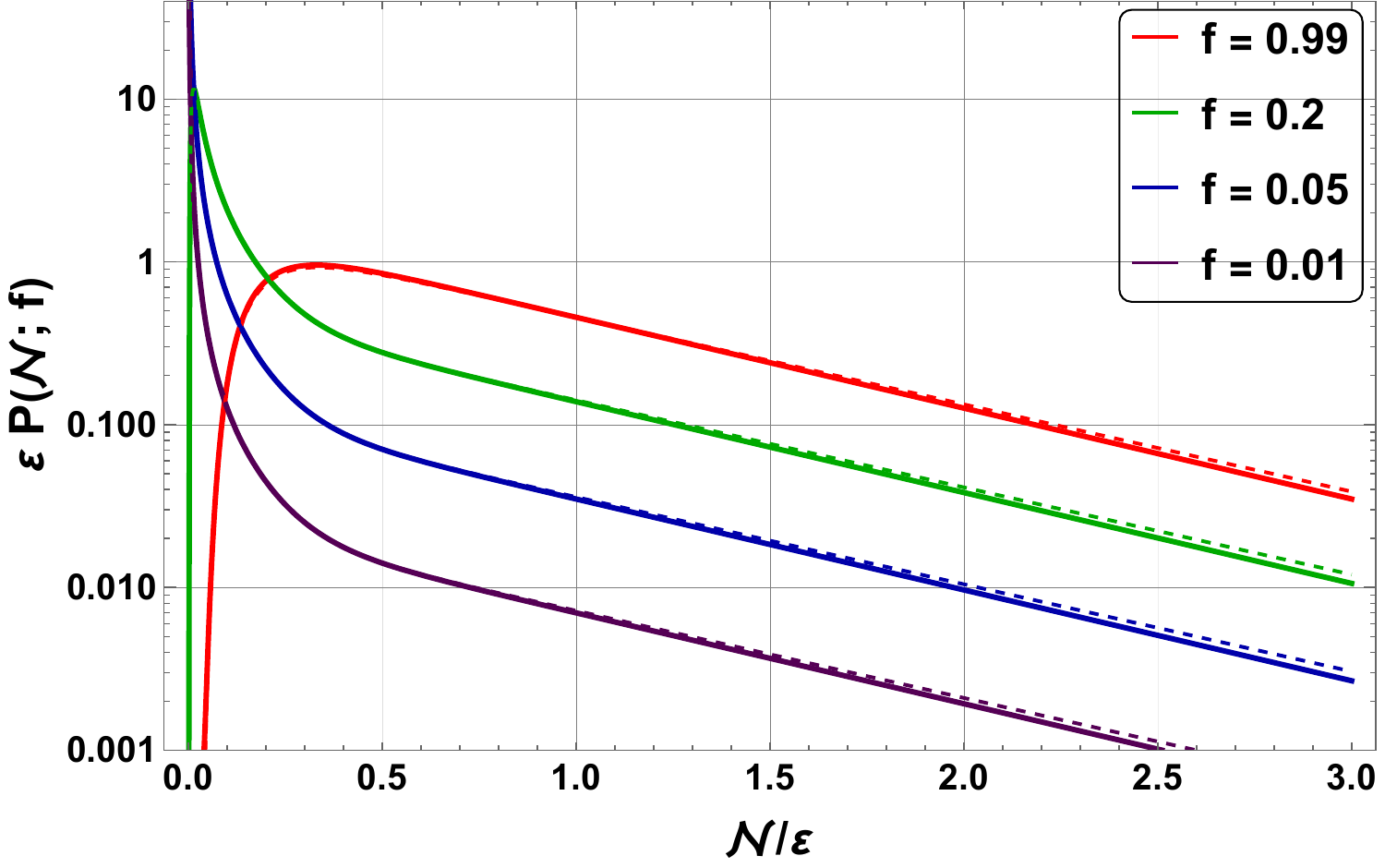}
\includegraphics[width=0.47\textwidth]{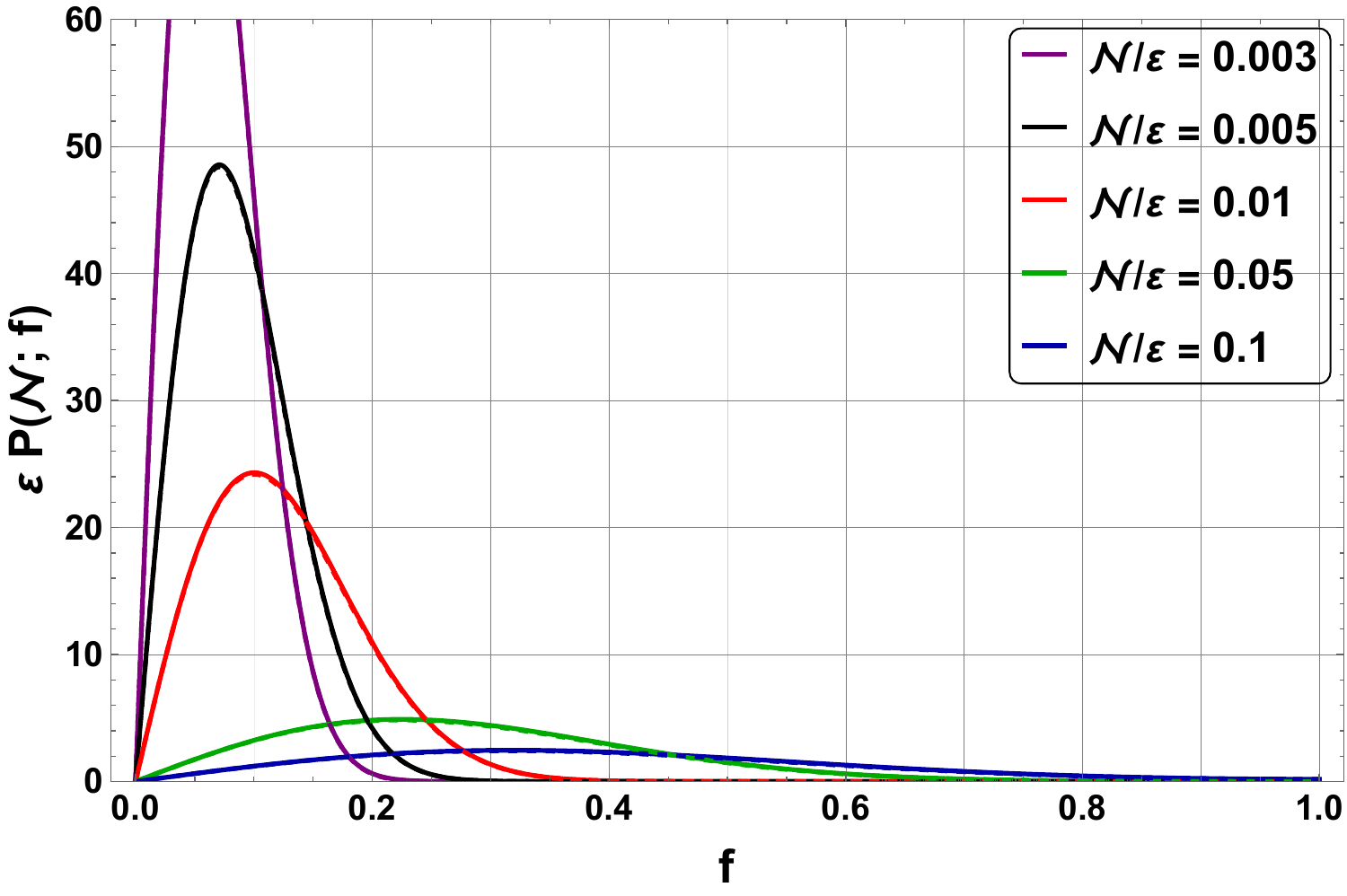}
\caption{The rescaled PDF, $\varepsilon \, P({\cal N}; f)$ from Eq.~\eqref{eq:CD_diff_dom_PDF_NW_red_final},  for constant-drift inflation in the narrow-well approximation, to linear order in $\alpha$ with $\alpha = 0.1$ (solid lines)  as a function of ${\cal N}/\varepsilon$ for different values of  $f$ (left panel) and as a function of $f$ for different values of $\cal N/\varepsilon$ (right panel). The dashed lines show the rescaled PDF for the drift-free diffusion case ($\alpha=0$ in Eq.~\eqref{eq:CD_diff_dom_PDF_NW_red_final}).}
\label{fig:PDF_CD_NW}
\end{center}
\end{figure}

In the narrow-well limit, $\alpha \ll 1$,  the expression for $\xi_n$ appearing in the solution  for ${\cal Z}_n$, in  Eq.~(\ref{eq:CD_diff_dom_Zn_sol_new}), can be written as
\beq
\xi_n^{\rm NW} = \f{\pi}{2} - \delta_n \, ; \quad {\rm with} ~ 0 \leq \delta_n \ll 1 \, ,
\label{eq:CD_diff_dom_varepsilon_NW}
\eeq
 where, recall, $n \geq 0$ is a non-negative  integer. Eq.~\eqref{eq:CD_diff_dom_varepsilon_NW} reduces Eq.~(\ref{eq:CD_diff_dom_tan_cond_gen}) to
$$ \tan{\l( \f{\pi}{2} - \delta_n \r)} = \l[ (n+1) \, \pi -  \l(\f{\pi}{2} - \delta_n \r)\r] \l( \f{2}{\alpha } \r)  \, .$$
Recalling $\delta_n \ll 1$, expanding the $\tan$ expression to leading order in $\delta_n$, this simplifies to the linearised solution 
$$\delta_n = \f{\alpha }{(2n+1)\pi} \l( 1  - \f{\delta_n}{(2n+1)\pi/2}\r) + {\cal O}(\delta_n^2) \, ,$$
which, at leading order in $\alpha$, yields
\beq
\delta_n =  \f{\alpha }{(2n+1)\pi} + {\cal O}\l(\f{\alpha }{(2n+1)\pi}\r)^2\,, 
\label{eq:CD_NW_deltan_final}
\eeq
a result which is valid when $\delta_n \ll 1$ hence  
$\alpha/[(2n+1)\pi] \ll 1$.
When that limit is satisfied we have 
\beq
\xi_n^{\rm NW} = \f{\pi}{2}  -  \f{\alpha }{(2n+1)\pi}\, , \quad  \Rightarrow \quad {\cal Z}_n^{\rm NW} = (2n+1)\f{\pi}{2}  + \f{\alpha}{(2n+1)\pi} + {\cal O}\l(\f{\alpha}{(2n+1)\pi}\r)^2\, ,
\label{eq:epsilon_Z_NW}
\eeq
leading to the final expression(s) for the quantised exponents $\Lambda_n$,   Eq.~\eqref{eq:CD_diff_dom_Lambda_n_gen}, expanded to leading order in $\alpha$:
\beq
\Lambda_n^{\rm NW} =  \Lambda_n^{\rm Free} + \f{\alpha}{2 \varepsilon}  = \Lambda_n^{\rm Free} \l[ 1 +  \f{4 \alpha}{(2n+1)^2\pi^2} \r] + {\cal O}(\alpha^2) \, , 
\label{eq:CD_diff_dom_NW_Lambda_n}
\eeq 
where $\Lambda_n^{\rm Free}$ is the exact quantized exponents for the case of drift-free quantum diffusion, given by Eq.~(\ref{eq:FPE_free_Lambda}).
From Eq.~(\ref{eq:CD_diff_dom_NW_Lambda_n}), it is easy to see that $\Lambda_n^{\rm NW} \to \Lambda_n^{\rm Free}$ as $\alpha \to 0$ (or equivalently, $D_\phi \to 0$), as expected. Therefore, the narrow-well limit of constant-drift inflation results in a small enhancement in the exponents $\Lambda_n$, compared to the drift-free case. Substituting ${\cal Z}_n^{\rm NW}$ from 
Eq.~\eqref{eq:epsilon_Z_NW}, the approximate eigenfunctions in Eq.~(\ref{eq:FPE_eigen_CD_Abs_sol}),  become 
\beq
\Psi_n^{\rm NW}(f) \approx  A_n^{\rm NW}  \, \exp{ \left( \f{\alpha}{2}  f \right)} \,  \sin{ \l\{ (2n+1)\f{\pi}{2} \l[ 1 + \f{2 \alpha}{(2n+1)^2\pi^2} \r] f \r\} }    \, .
\label{eq:CD_diff_dom_eigen_fun_NW}
\eeq
The coefficient $A_n^{\rm NW}$ can be determined by using the orthonormality condition in Eq.~(\ref{eq:Psi_n_othonormal_correct}) in the narrow-well limit $\alpha \ll 1$.  Keeping terms up to linear order in $\alpha$  (see App.~\ref{app:CD_calculation_NW}) we find
\beq
A_n^{\rm NW} =    \sqrt{\f{2}{w_0\Delta\phi_{\rm well}}}  \l[ 1 - \f{\alpha}{(2n+1)^2\pi^2} \r]+ {\cal O}(\alpha^2)\,  .
\label{eq:CD_NW_An}
\eeq
Returning to Eq.~\eqref{eq:CD_diff_dom_eigen_fun_NW},
including Eq.~\eqref{eq:CD_NW_An}, and expanding to linear order in $\alpha$ we obtain the final expression for the  eigenfunction in the narrow-well approximation
\ber
\Psi^{\rm NW}_n(f) & = & 
\sqrt{\f{2}{w_0\Delta\phi_{\rm well}}}  \,
   \Biggl(  
            \sin{\l[ (2n+1)\f{\pi}{2} f \r]}    \nonumber \\
   && + \f{\alpha}{2}  \, 
             \biggl\{ 
                    \l[ f  - \f{2}{(2n+1)^2\pi^2}\r] \, \sin{\l[ (2n+1)\f{\pi}{2}  f \r]} \nonumber \\
                &&   + \, \f{2f}{(2n+1)\pi} \,   \cos{\l[ (2n+1)\f{\pi}{2}  f \r]} 
                \biggr\}           
    \Biggr) + {\cal O}(\alpha^2) \, .
\label{eq:CD_diff_dom_eigen_fun_NW_final_reduced_alpha}
\eer
Inserting Eq.~(\ref{eq:CD_diff_dom_eigen_fun_NW_final_reduced_alpha}) in  Eq.~(\ref{eq:To_B_f}), and keeping terms up to linear order in $\alpha$, we find (see App.~\ref{app:CD_calculation_NW})
\beq
{\cal B}^{\rm NW} = \f{\Delta\phi_{\rm well}^2}{2\varepsilon} + {\cal O}(\alpha^2)\, ,
\label{eq:CD_NW_B_int2}
\eeq
which is the same as is the exact drift-free diffusion case, Eq.~\eqref{eq:B_free_final}, as $\alpha \to 0$.  
Finally,  the PDF from Eq.~(\ref{eq:PDF_N_f}) in the narrow-well limit, keeping terms  up to linear order in $\alpha$, takes the form  
\ber
P({\cal N}; f) &=& \l(\f{1}{\varepsilon}\r) \f{\pi}{2} \,   \sum_{n=0}^{\infty} \, (2n+1)  
      \, 
           \sin{\l[ (2n+1)\f{\pi}{2} f \r]}    \nonumber \\
                &+& \f{\alpha f}{2}  ~ 
                \biggl\{ 
                     \, (1+2n) \sin{\l[ (2n+1)\f{\pi}{2}  f \r]} + \, \f{2}{\pi} \,  \cos{\l[ (2n+1)\f{\pi}{2}  f \r]} 
               \, \biggr\}           
       \nonumber \\
& \times &   \exp{\l\{- \l[ 1 + \f{4\alpha}{(2n+1)^2 \pi^2} \r] \, \l( 2n+1 \r)^2 \, \f{\pi^2}{8} \, \f{1}{\varepsilon} \, {\cal N} \r\} } + {\cal O}(\alpha^2) \, .
\label{eq:CD_diff_dom_PDF_NW_red_final}
\eer
The rescaled PDF $\varepsilon \, P({\cal N}; f)$ is plotted as a function of ${\cal N}/\varepsilon $ and $f$ in Fig.~\ref{fig:PDF_CD_NW}.
 Since $\alpha \ll 1$ in the narrow-well limit, the PDF of constant-drift inflation is very close to that of the free-diffusion case for ${\cal N}/\varepsilon \lesssim 1$. However, in the tail where ${\cal N}/\varepsilon \gtrsim 1$, the PDF deviates appreciably from that of the free-diffusion case. As expected Eq.~\eqref{eq:CD_diff_dom_PDF_NW_red_final}
reproduces the PDF of the drift-free case Eq.~\eqref{eq:FPE_free_PDF_final} in the limit $\alpha \to 0$.  For the parameter range considered in the narrow-well regime, summing $\mathcal{O}(10^3)$ modes is sufficient to achieve numerical convergence of the PDF.
 
\begin{figure}[!t]
\begin{center}
\includegraphics[width=0.7\textwidth]{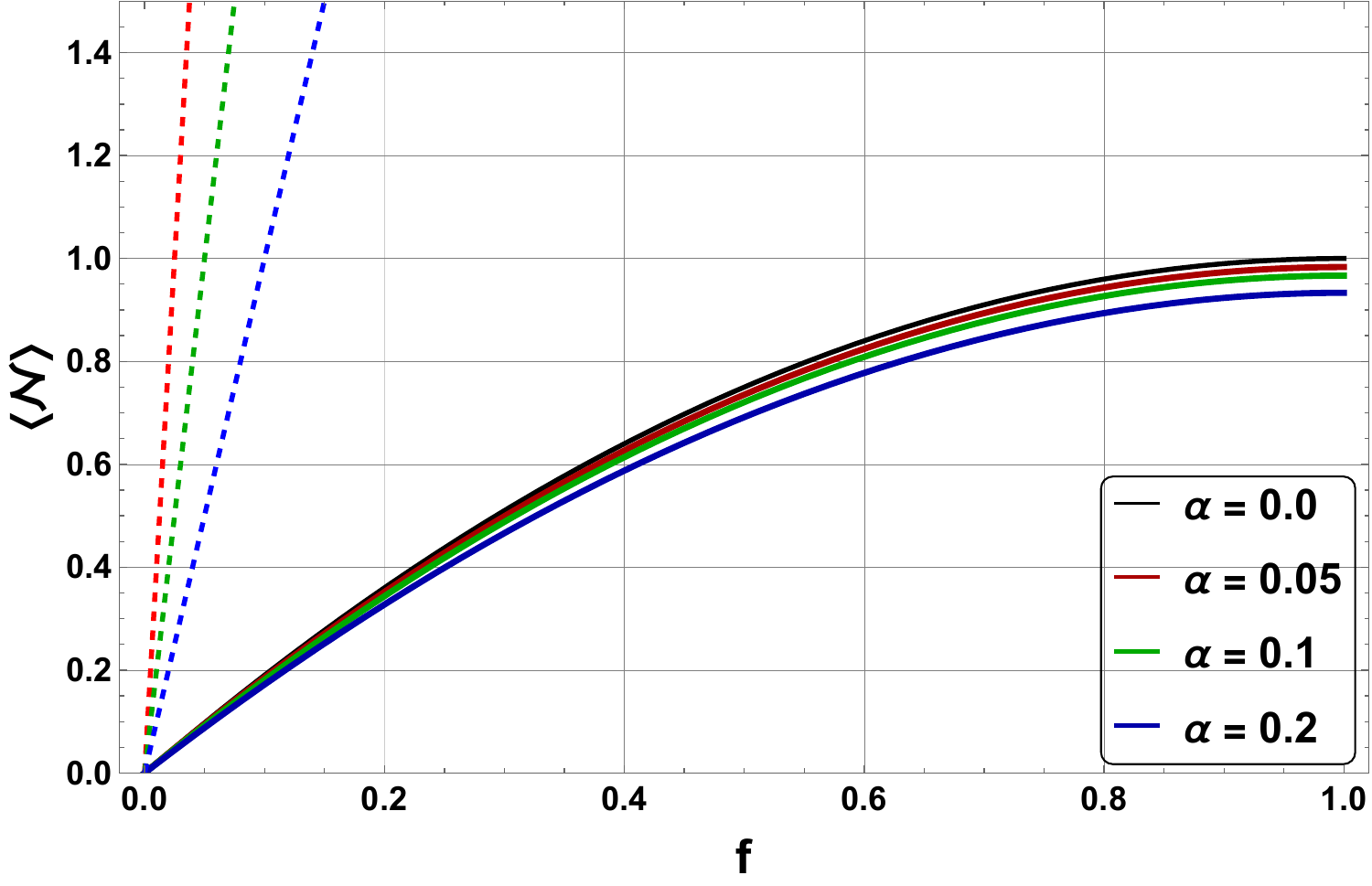}
\caption{The expectation value of the 
stochastic number of e-folds,  $\langle {\cal N}\rangle$, for constant-drift inflation in the narrow-well limit, given by Eq.~\eqref{eq:CD_NW_PDF_N_avg}, as a function of $f$
for $\varepsilon=1$ and $\alpha  = 0.05,\,0.1$ and $0.2$ (in red, green and  blue, respectively). The dashed lines show the classical number of e-folds, as defined in Eq.~\eqref{eq:N_cl_CD}. The black line shows $\langle {\cal N}\rangle$ for the free diffusion case with $\alpha=0$, as given by Eq.~\eqref{eq:N_moment1_Free1}.
}
\label{fig:CD_NW_Navg_SI_Ncl}
\end{center}
\end{figure}

To obtain the PDF $P\l(\delta{\cal N};\,f\r)$ in Eq.~\eqref{eq:PDF_delta_N_f},  we  compute $\langle {\cal N} \rangle$, by inserting  Eq.~\eqref{eq:CD_diff_dom_NW_Lambda_n}, 
Eq.~\eqref{eq:CD_diff_dom_eigen_fun_NW_final_reduced_alpha} and Eq.~\eqref{eq:CD_NW_B_int2} in Eq.~\eqref{eq:N_avg_f}. Keeping terms up to linear order in $\alpha$, and using the {\em sine and cosine sum formulae},  Eqs.~\eqref{eq:Sum_sin_odd_3A},~\eqref{eq:Sum_cos_even_4A},~\eqref{eq:Sum_sin_odd_5A} in App.~\ref{app:sums}, we find (see App.~\ref{app:CD_calculation_NW} for details):
\beq
\langle {\cal N} \rangle = \varepsilon \, f \l[ (2-f) + \f{\alpha}{3} \l( -3 + 3 \, f - f^2 \r) \r] + {\cal O}(\alpha^2) \,.
\label{eq:CD_NW_PDF_N_avg}
\eeq
This reduces to the expression for the drift-free diffusion case in Eq.~(\ref{eq:N_moment1_Free1}) as $\alpha \to 0$, as expected. Furthermore, Fig.~\ref{fig:CD_NW_Navg_SI_Ncl} shows that the stochastic average  number of e-folds is much smaller than the classical number of e-folds given in Eq.~(\ref{eq:N_cl_CD}), independent of the value of $\alpha$. This is expected, because the narrow-well limit of constant-drift inflation is closer to the free-diffusion case,  whereas the classical number of e-folds, being inversely proportional to $\alpha$, tends to infinity in the absence of classical drift, whilst stochastic effects produce a finite average number of first-passage e-folds.
The PDF, $P\l(\delta{\cal N};\,f\r)$, follows directly from Eq.~\eqref{eq:CD_diff_dom_PDF_NW_red_final} upon the replacement $ {\cal N}  \rightarrow \delta{\cal N} + \langle {\cal N} \rangle$. 

\subsubsection{Broad-well limit}
\label{sec:ET_CD_BW}
\begin{figure}[!t]
\begin{center}
\includegraphics[width=0.487\textwidth]{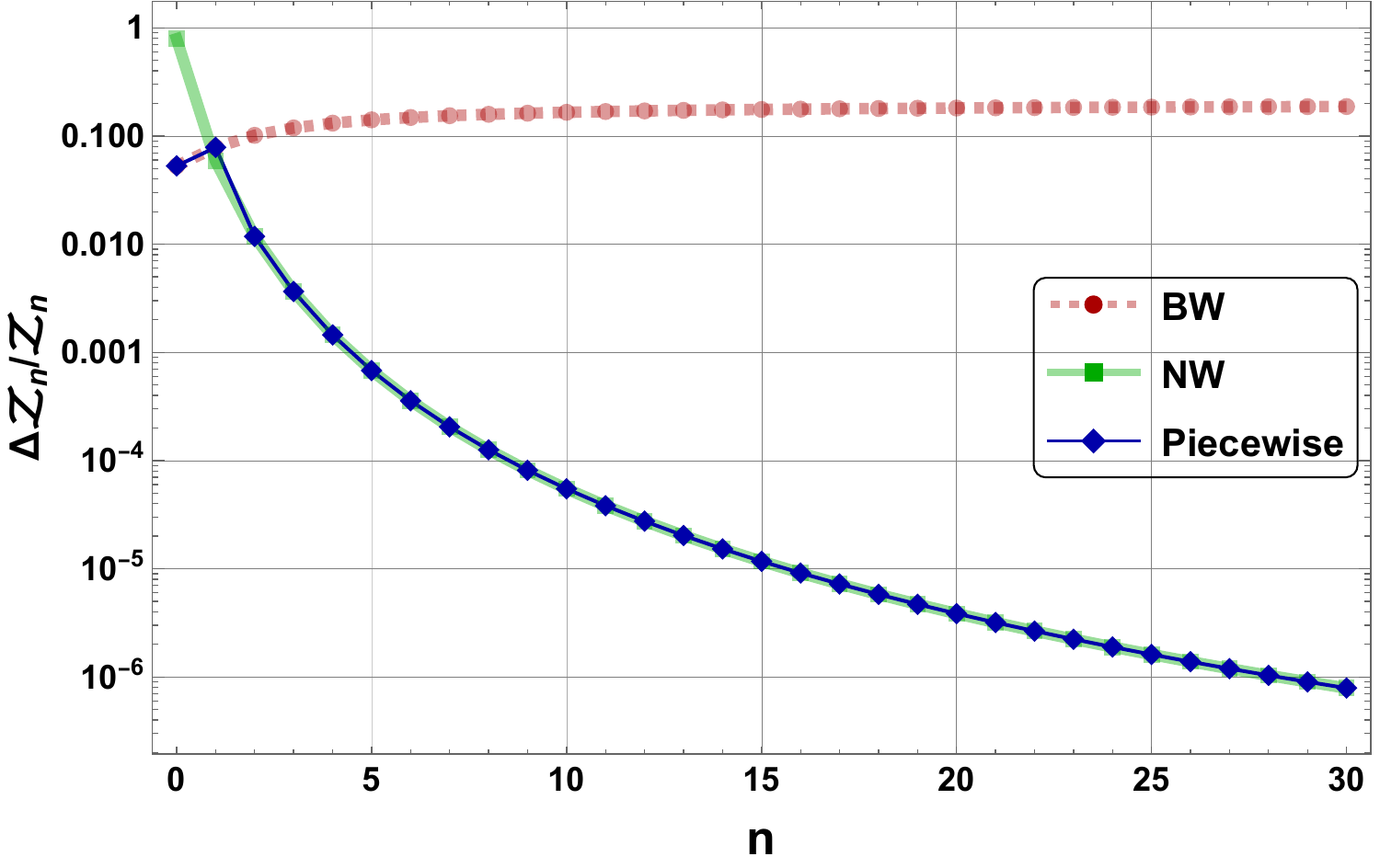}
\includegraphics[width=0.487\textwidth]{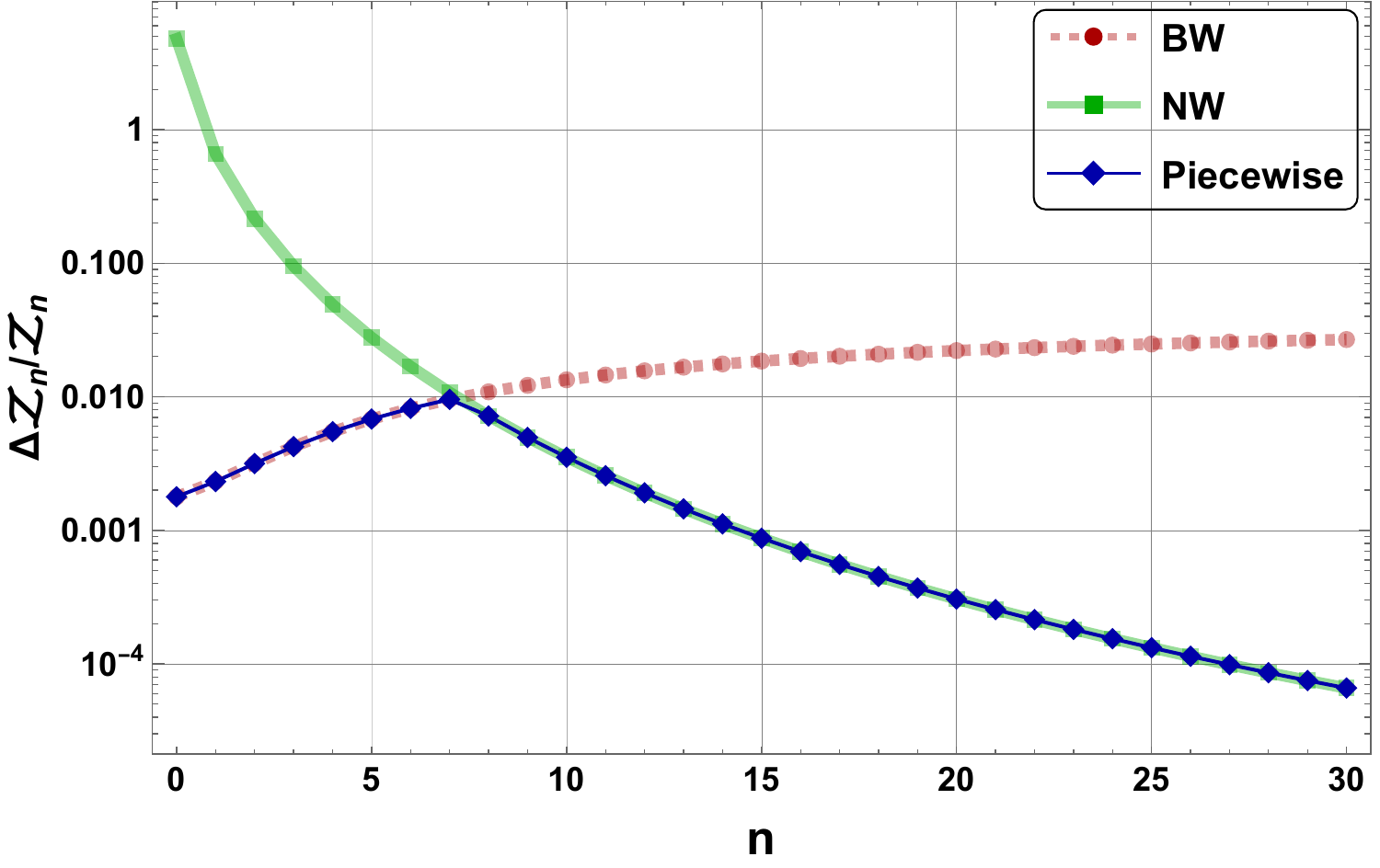}
\caption{The fractional error in ${\cal Z}_n$, defined through Eq.~\eqref{eq:Zn_Err}, as a function of $n$ for $\alpha = 10$ (left panel) and $\alpha = 50$ (right panel). The green and red lines show the fractional error obtained using the narrow-well approximation, Eq.~\eqref{eq:epsilon_Z_NW}, and broad-well approximation, Eq.~\eqref{eq:epsilon_Z_BW}, respectively, while the blue line corresponds to the piecewise construction, Eq.~\eqref{zncases}, with $n_c$ given by Eq.~\eqref{eq:nc_def}.}
\label{fig:CD_Zn}
\end{center}
\end{figure}

 The broad-well limit,  $\alpha \gg 1$, corresponds to $\xi_n \to 0$ in Eq.~\eqref{eq:CD_diff_dom_Zn_sol_new}, and Eq.~(\ref{eq:CD_diff_dom_tan_cond_gen}) reduces to
$$\tan{\l(\xi_n\r)}  \simeq \xi_n = \f{2}{\alpha} \l[ (n+1)\pi - \xi_n\r] + {\cal O}(\xi_n^3)
,$$
which, using Eq.~(\ref{eq:CD_diff_dom_Zn_sol_new}), yields to leading order in $1/\alpha$
\ber
\xi_n^{\rm BW} &=& \f{2}{\alpha} \l(n+1\r)\pi +{\cal O}\l(\f{2}{\alpha} \l(n+1\r)\pi\r)^2 \,, \nonumber \\  \quad &\Rightarrow & \quad {\cal Z}_n^{\rm BW} = (n+1) \pi \l( 1 - \f{2}{\alpha} \r) +{\cal O}\l(\f{2}{\alpha} \l(n+1\r)\pi\r)^2 \, ,
\label{eq:epsilon_Z_BW}
\eer
where $n \geq 0$ is a non-negative  integer. Therefore in this regime the quantised exponents in Eq.~\eqref{eq:CD_diff_dom_Lambda_n_gen} can be approximated as 
\beq
 \Lambda_n^{\rm BW} = \f{\alpha^2}{8\,\varepsilon} \,  \l[ 1 + \f{4\pi^2 \, (n+1)^2}{\alpha^2}\r] + {\cal O}\l(\f{2(1+n)^2\pi^2}{\alpha \, \varepsilon}\r)
 \,,
\label{eq:CD_diff_dom_Lambda_n_BW}
\eeq 
Clearly, Eq.~\eqref{eq:epsilon_Z_BW} only holds as $\xi_n \to 0$ or 
$2\pi (n+1)/\alpha \ll 1$. We immediately see a potential issue for all but the very largest values of $\alpha$, since the PDF defined in Eq.~\eqref{eq:PDF_N_f} involves an infinite sum over $n$. Although one might think the exponential tail of the distribution is highly suppressed for large values of $n$, this is not the case when we are considering correspondingly small values of ${\cal N}/\varepsilon$. In those cases we are faced with the fact that we are no longer in the regime
$2\pi (n+1)/\alpha \ll 1$,
hence the solution for $\Lambda_n^{\rm BW}$, Eq.~\eqref{eq:CD_diff_dom_Lambda_n_BW}, is no longer valid. In reality the system transitions from a period where Eq.~\eqref{eq:CD_diff_dom_Lambda_n_BW} is the correct eigenvalue, to one where Eq.~\eqref{eq:CD_diff_dom_NW_Lambda_n} is the correct behaviour, as it corresponds to the solution in the regime
$\alpha/[(2n+1)\pi] \ll 1$.  By using the appropriate approximate values of ${\cal Z}_n$ in the limits $n \leq n_c$ and $n > n_c$, where
\beq
n_c =   {\rm IntegerPart}\l[ \f{1}{2}\l(\f{\alpha}{\pi} -1\r)\r]\, ,
\label{eq:nc_def}
\eeq
we formulate a  piecewise construction of  ${\cal Z}_n$,  
\begin{equation}
{\cal Z}_n^{\rm Piecewise} =
\begin{cases}
(n+1) \pi \l(1-\f{2}{\alpha}\r) &  {\rm for} ~ n \leq n_c \,, \\ 
(2n+1)  \f{\pi}{2} \l(1+\f{2\alpha}{(2n+1)^2\pi^2}\r) ~
& {\rm for} ~  n > n_c \,,\end{cases}
\label{zncases}
\end{equation} 
which can be used to determine the broad-well eigenfunctions, Eq.~\eqref{eq:FPE_eigen_CD_Abs_sol}, analytically. The fractional error in the true numerical value of ${\cal Z}_n$, Eq.~\eqref{eq:FPE_CD_Zn_quant}, when compared to the piecewise expression in Eq.~\eqref{zncases}  
is defined as 
\beq
\f{\Delta{\cal Z}_n}{{\cal Z}_n} = \l|\f{{\cal Z}_n-{\cal Z}_n^{\rm Piecewise}}{{\cal Z}_n}\r| \, 
\label{eq:Zn_Err}
\eeq
and is shown in Fig.~\ref{fig:CD_Zn} for two representative values of $\alpha$. 
We see that the piecewise solutions agree well with the full numerical solutions in the relevant regimes of $n$, with fractional errors below 10\% for $\alpha=10$, and below 1\% for $\alpha=50$.

The coefficient $A_n^{\rm BW}$  in the broad-well limit, $\alpha \gg 1$, can be approximated from Eq.~\eqref{An1} as 
\beq
  A_n^{\rm BW} =  \l(\f{2}{w_0\Delta\phi_{\rm well}}\r)^{1/2}  
 \,.
\label{AnBW}
\eeq

 Turning to ${\cal B}^{\rm BW}$ in Eq.~\eqref{eq:To_B_f1}, we note that it not only involves an infinite sum over $n$,  which requires  us to accommodate the changing behaviour of ${\cal Z}_n$ given in Eq.~\eqref{zncases}, but it also involves taking the subsequent limit $f\to 0$. Hence, a full numerical determination of ${\cal B}^{\rm BW}$ becomes challenging. Therefore, we obtain ${\cal B}^{\rm BW}$ analytically, by making use of the large $\alpha$ expansion of ${\cal Z}_n$ given in Eq.~\eqref{eq:epsilon_Z_BW}. Keeping terms up to leading order in $\alpha$, we find  
\beq
{\cal B}^{\rm BW} \simeq   \f{\Delta\phi_{\rm well}^2}{2\,\varepsilon} \, .
\label{eq:CD_BW_B_int2-Ed1}
\eeq
Using  Eq.~\eqref{eq:CD_BW_B_int2-Ed1},  in the broad-well regime the PDF, Eq.~\eqref{eq:PDF_N_f_CD_Gen}, becomes approximated by  
\beq
P({\cal N};\,f)\simeq \l(\f{1}{\varepsilon}\r) e^{{\l(\f{\alpha}{2}\,f\r)}}\,
\sum_{n=0}^{\infty} \, {\cal Z}_n \, \sin({\cal Z}_n f) \, e^{-\f{1}{2}\l({\cal Z}_n^2 + \f{\alpha^2}{4}\r) \f{\cal N}{\varepsilon}} \,.
\label{PdfBWnew}
\eeq

\begin{figure}[!t]
\begin{center}
\includegraphics[width=0.487\textwidth]{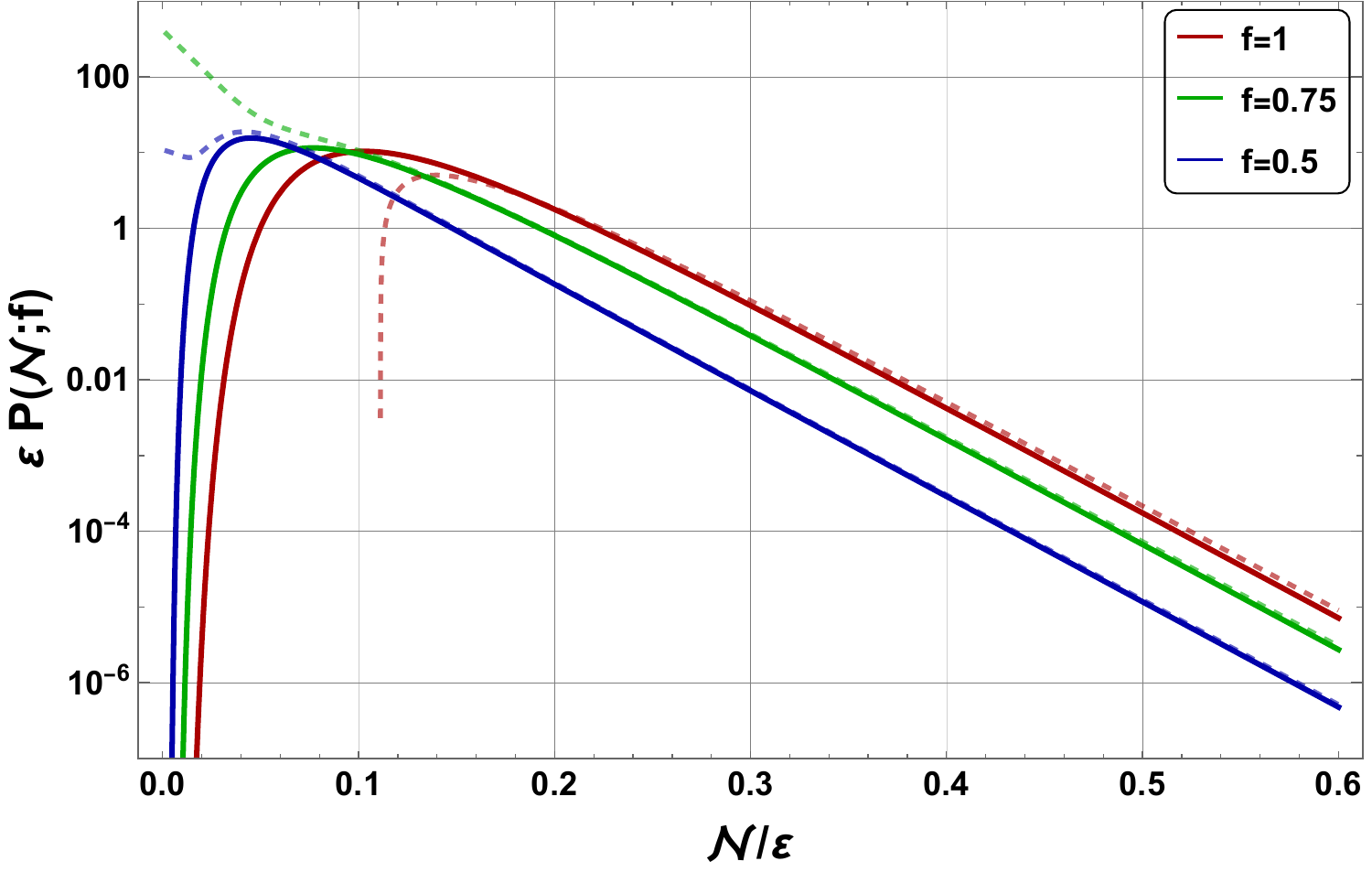}
\includegraphics[width=0.487\textwidth]{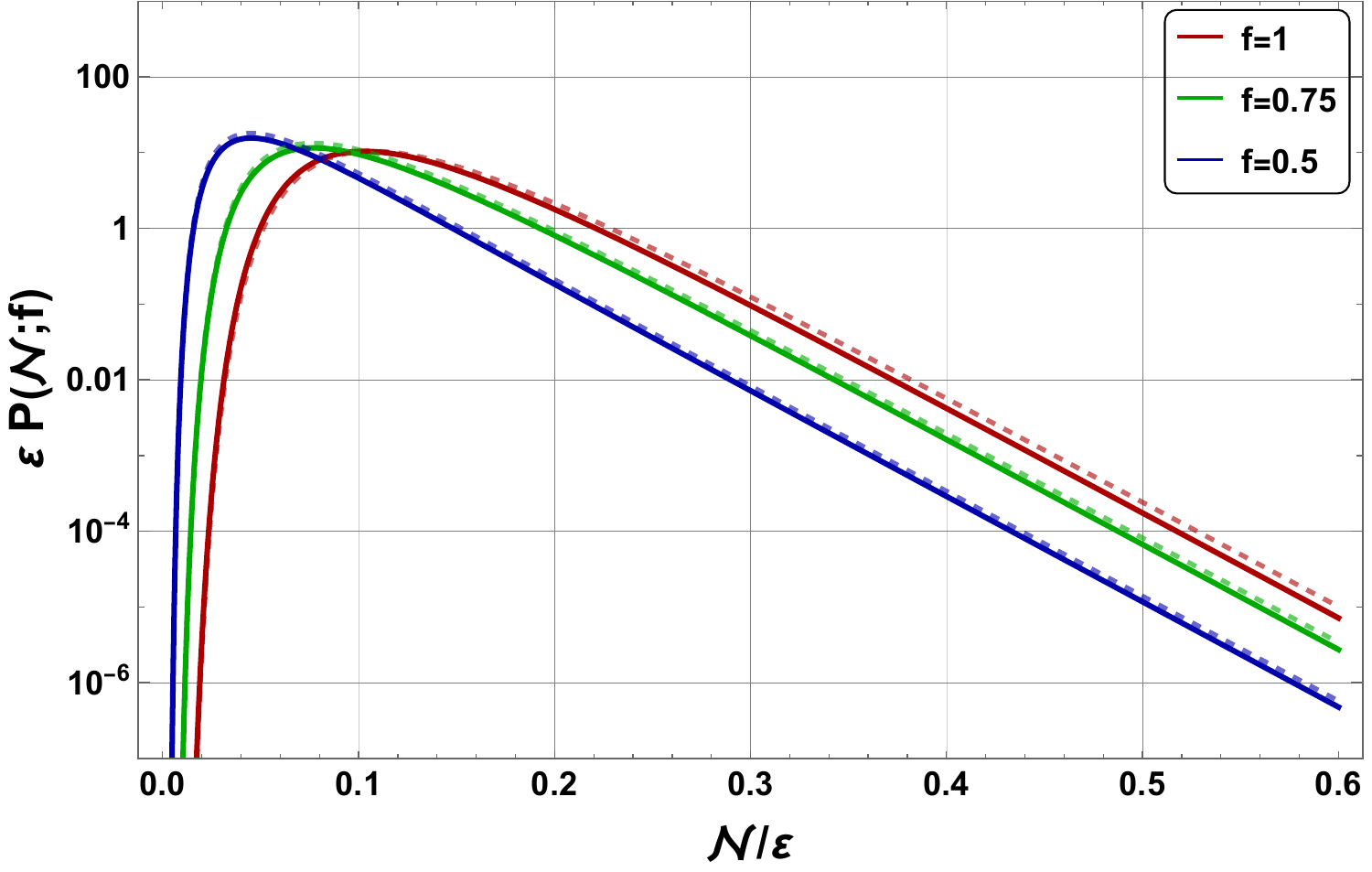}
\caption{The rescaled PDF, $\varepsilon\,P({\cal N};\,f)$,  Eq.~\eqref{PdfBWnew}, determined numerically as a function of ${\cal N}/\varepsilon$ for constant-drift inflation in the broad-well limit for $\alpha = 15$ (solid lines). The red, green and blue lines correspond to $f=1, \,0.75,$ and $0.5$ respectively. Dashed lines in the left panel represent the corresponding PDFs determined by using the piecewise construction of ${\cal Z}_n$ given in Eq.~\eqref{zncases}. While, dashed lines in the right panel correspond  to the PDF determined using the large $\alpha$ expansion of ${\cal Z}_n$ given in Eq.~\eqref{eq:PDF_CD_BW_elliptic}.}
\label{fig:PDF_CD_BW_Num_Piecewise}
\end{center}
\end{figure}

 The PDF in Eq.~\eqref{PdfBWnew} can be obtained numerically by using the piecewise analytical construction of ${\cal Z}_n$ given in Eq.~\eqref{zncases}. Since the PDF is expected to feature a peak  for small ${\cal N}/\varepsilon$, typically one needs to carry out the summation in Eq.~\eqref{PdfBWnew} over $n$ from $n=0$, to some large enough $n=n_{\rm max} \gtrsim n_c$.   Unfortunately, for the piecewise construction of ${\cal Z}_n$ in Eq.~\eqref{zncases}, it turns out that the switch in the eigenspectrum at $n=n_c$, can introduce artificial features in the summation in Eq.~\eqref{PdfBWnew}, particularly for larger values of $f$. This leads to an inaccurate determination of the peak of the PDF\footnote{This occurs because the PDF is obtained from a  superposition of many eigenfunctions, and is therefore highly sensitive not only to the accuracy of the individual eigenvalues, but also to the coherence of the spectrum as a function of $n$.  This coherence breaks down due to the sudden switch in ${\cal Z}_n$ at $n=n_c$, and  the piecewise asymptotic construction does not preserve the approximate weighted orthogonality of the eigenfunctions, Eq.~\eqref{eq:Psi_n_othonormal_f}. This behaviour is reminiscent of the Gibbs phenomenon, which can produce reconstruction artefacts in Fourier and other spectral analyses~\cite{boyd2001chebyshev}. } at smaller values of ${\cal N}/\varepsilon$. 
Therefore, although Eq.~\eqref{zncases} provides a good analytical approximation for determining  the eigenvalues and eigenfunctions, as well as the PDF at its tail, it becomes inadequate in  determining the peak of the PDF. In order to compute the PDF over the full range of ${\cal N}/\varepsilon$, we should ideally solve the transcendental Eq.~\eqref{eq:FPE_CD_Zn_quant} numerically. This is demonstrated in the left panel of Fig.~\ref{fig:PDF_CD_BW_Num_Piecewise} which shows that the PDF, determined using the piecewise construction, deviates significantly from its numerically determined value near the peak, whilst remaining a good approximation for the PDF near the tail of the distribution.

Given this issue with the piecewise construction of the PDF near the peak of the distribution we can ask the question, how well does the large $\alpha$ solution for ${\cal Z}_n$, namely Eq.~\eqref{eq:epsilon_Z_BW}, work in determining the PDF in Eq.~\eqref{PdfBWnew}? In that case, the infinite sum over $n$ can be performed yielding 
\beq
P({\cal N};\,f)\simeq -\f{\pi}{4}\l(\f{1}{\varepsilon}\r) \l(1-\f{2}{\alpha}\r) e^{{\l(\f{\alpha}{2}\,f\r)}}\,
e^{-\l(\f{\alpha^2\;\cal N}{8\;\varepsilon}\r)}
\;\vartheta_3'\l( \f{\pi}{2} \,\l(1-\f{2}{\alpha}\r)f \, ; ~ e^{-\f{\pi^2}{2} \, \f{{\cal N}}{\varepsilon}\l(1-\f{2}{\alpha}\r)^2} \r)\, ,
\label{eq:PDF_CD_BW_elliptic}
\eeq
where $\vartheta_3$ is the {\em Jacobi elliptic (theta) function} of the third kind~\cite{book_nist_gov}. Here $\vartheta_3'(X;Y) \equiv \f{\partial}{\partial X}\vartheta_3(X;Y)$.
  Interestingly, Eq.~\eqref{eq:PDF_CD_BW_elliptic} provides a more accurate approximation to the numerical PDF near the peak than that determined using the piecewise construction,  as demonstrated in the right panel  of Fig.~\ref{fig:PDF_CD_BW_Num_Piecewise}.   This is because it preserves the coherence of spectral structure for all values of $n$, despite providing a less accurate approximation to the individual eigenvalues for $n>n_c$.

 \begin{figure}[!t]
\begin{center}
\includegraphics[width=0.487\textwidth]{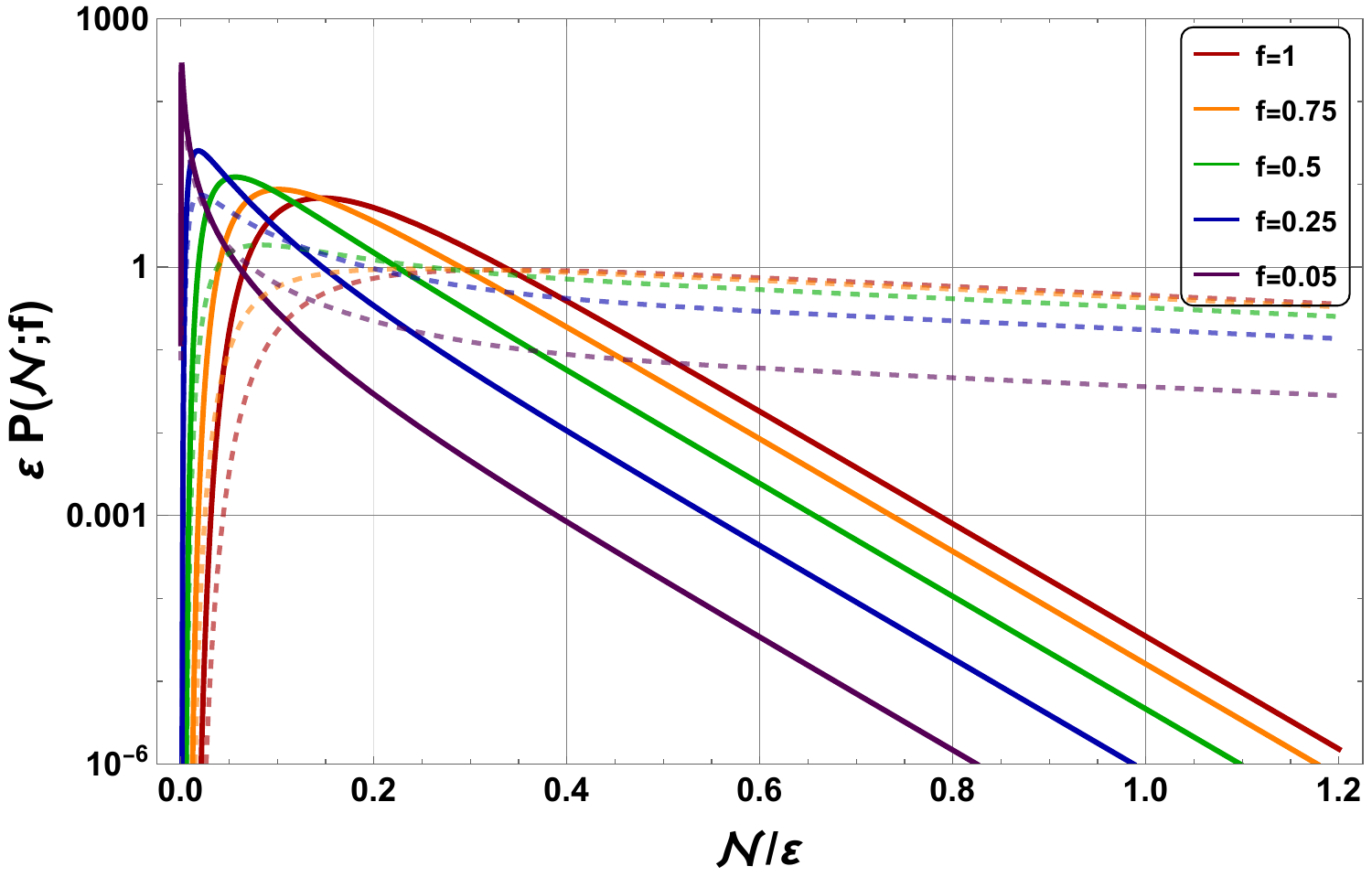}
\includegraphics[width=0.487\textwidth]{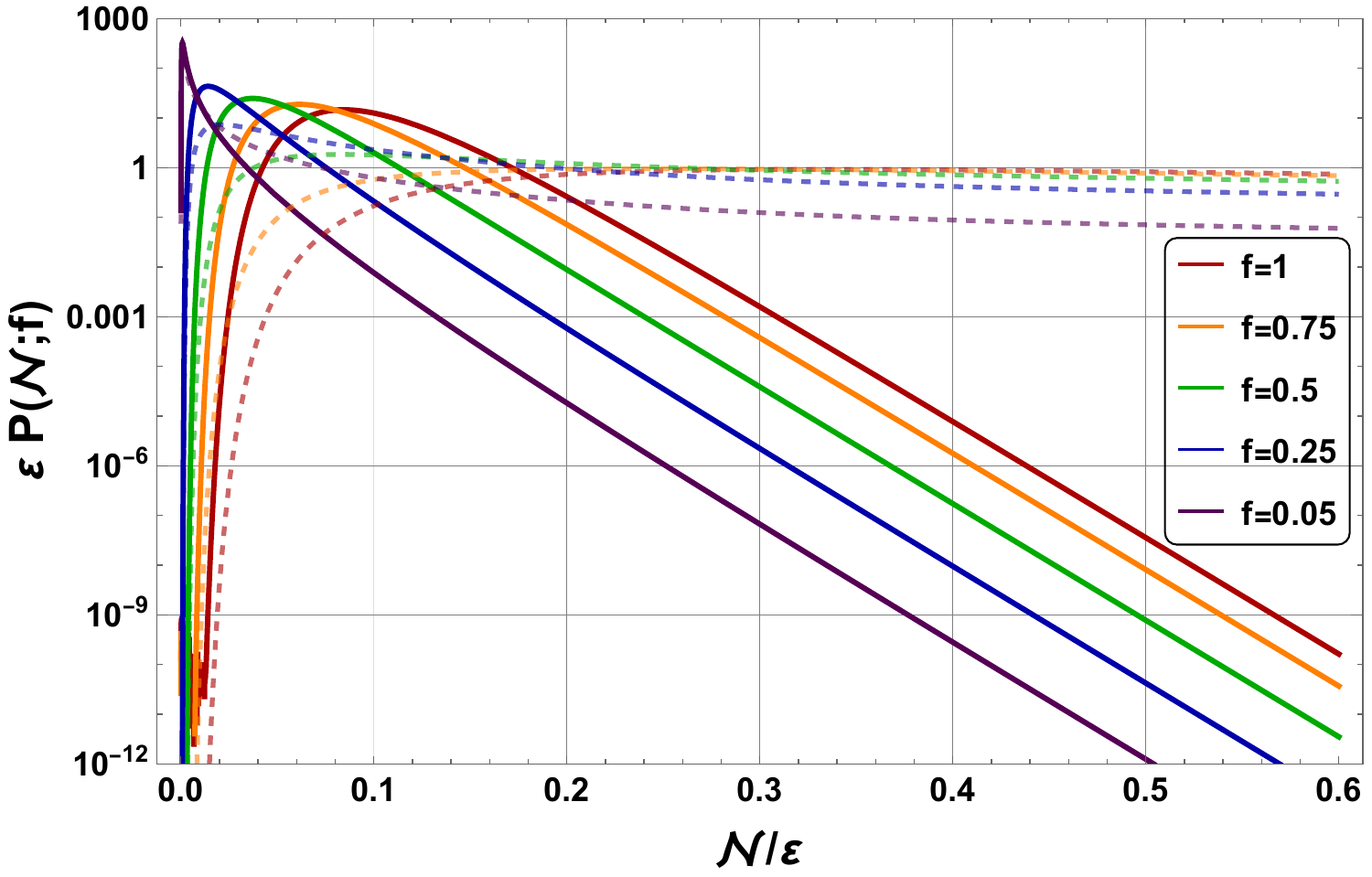}
\caption{The rescaled PDF, $\varepsilon\,P({\cal N};\,f)$,  Eq.~\eqref{PdfBWnew}, as a function of ${\cal N}/\varepsilon$ for constant-drift inflation in the broad-well limit for $\alpha = 10$ (left panel) and $20$ (right panel). The red, orange, green, blue and purple lines correspond to $f=1, 0.75, 0.5,\,0.25,$ and $0.05$ respectively, and the dashed lines show the corresponding rescaled PDFs for the free-diffusion case ($\alpha=0$).}
\label{fig:PDF_CD_BW_Free}
\end{center}
\end{figure}
\begin{figure}[!t]
\begin{center}
\includegraphics[width=0.487\textwidth]{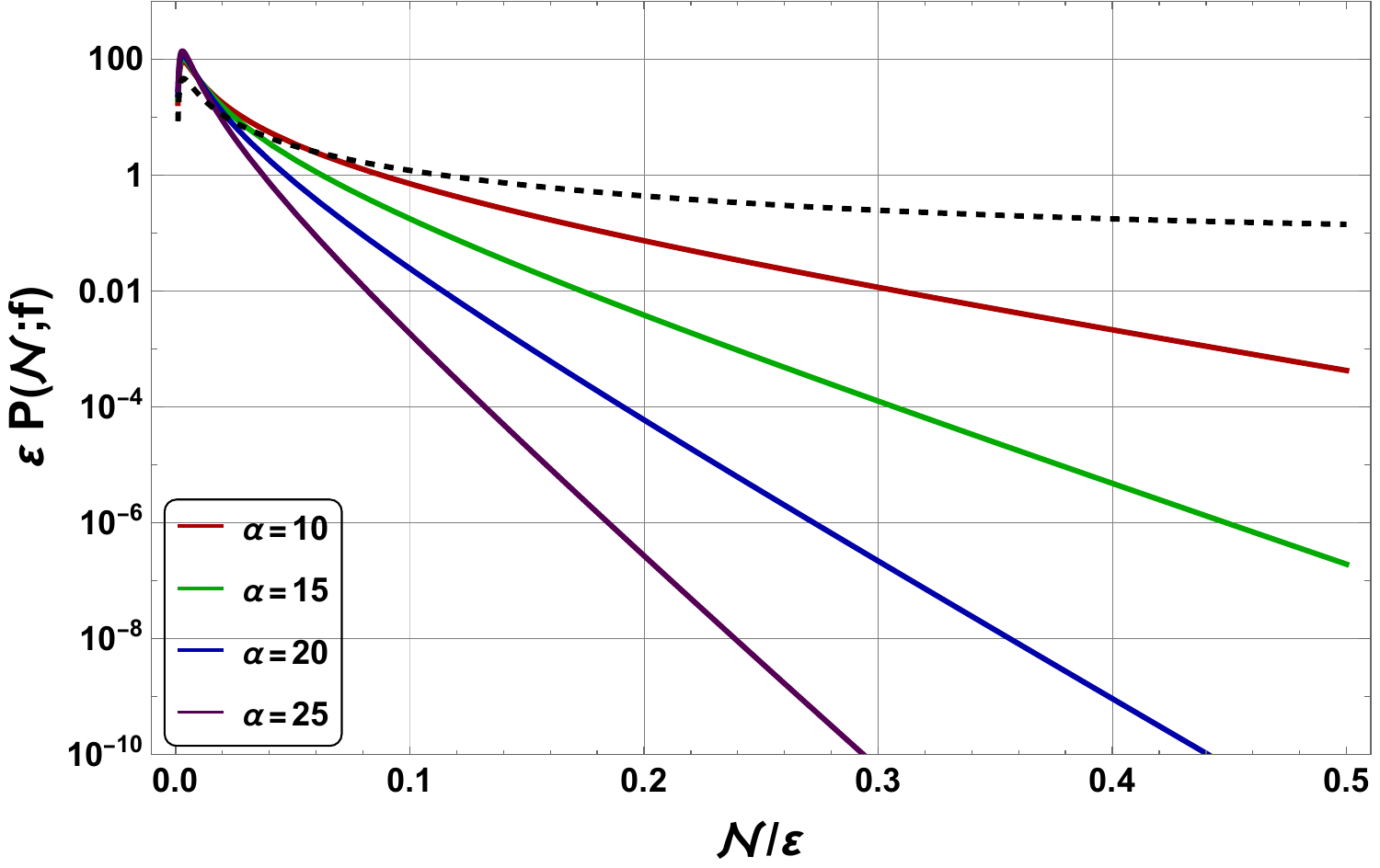}
\includegraphics[width=0.487\textwidth]{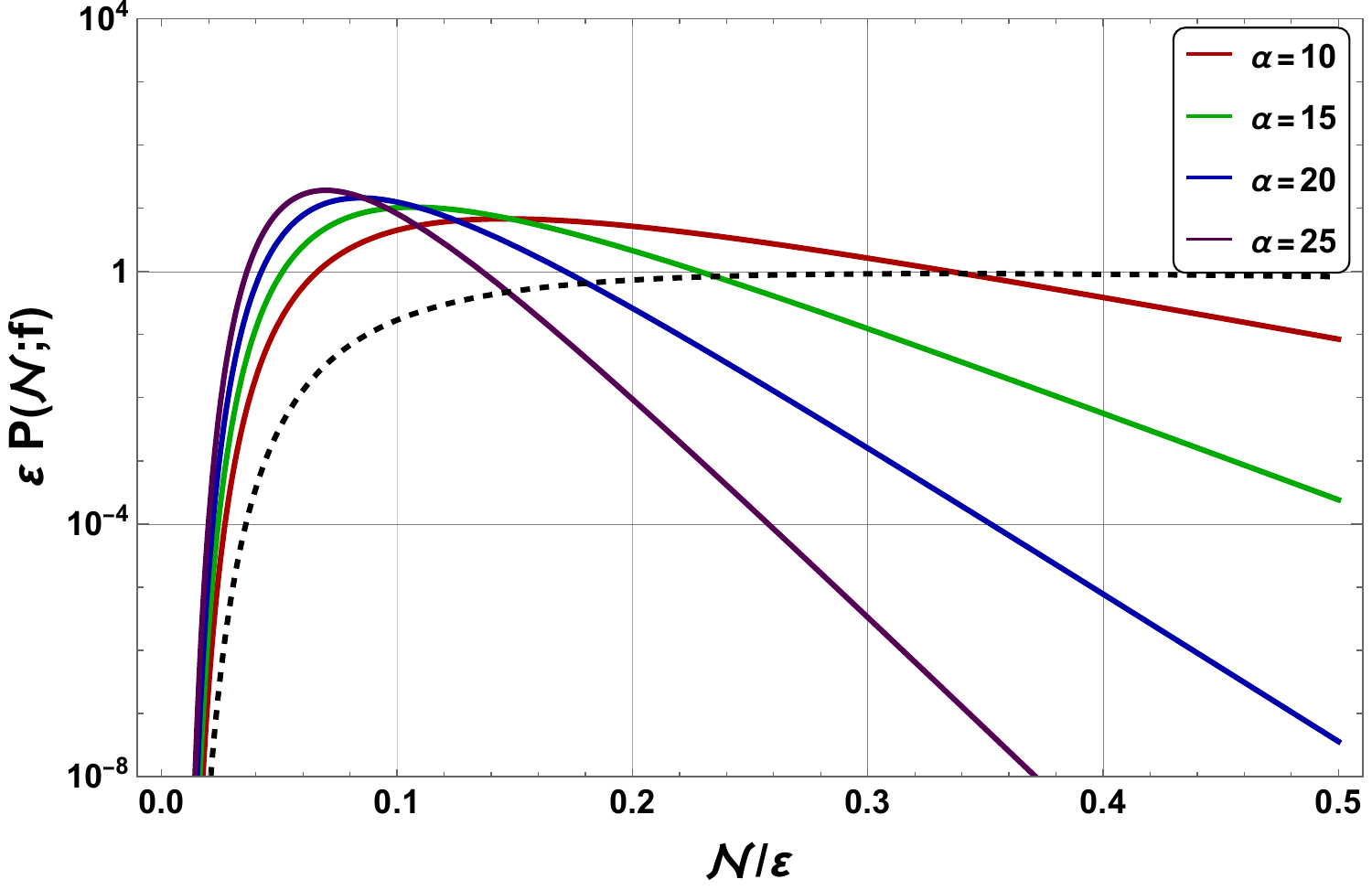}
\caption{The rescaled PDF, $\varepsilon\,P({\cal N};\,f)$,  Eq.~\eqref{PdfBWnew}, as a function of ${\cal N}/\varepsilon$ for constant-drift inflation in the broad-well approximation for $\alpha=10,\,15,\,20$ and $25$ (red, green, blue, and purple lines respectively) and $f = 0.1$ and $1$ (left and right panels respectively). The dashed black line shows the rescaled PDF for the free-diffusion case ($\alpha=0$).}
\label{fig:PDF_CD_BW}
\end{center}
\end{figure}

 Even though, Eq.~\eqref{eq:PDF_CD_BW_elliptic} provides a good fit to the PDF in the broad-well regime, in what follows we show the numerically determined\footnote{In the broad-well regime, $\mathcal{O}(10^3)$ modes are typically sufficient to achieve convergence over the parameter range considered here. However, summation of $\mathcal{O}(10^4
)$ modes was required for $f \leq 0.1$.} PDFs.  The rescaled PDF $\varepsilon\,P({\cal N};\,f)$, from Eq.~\eqref{PdfBWnew},  is shown in Figs.~\ref{fig:PDF_CD_BW_Free}~and~\ref{fig:PDF_CD_BW} as a function of ${\cal N}/\varepsilon$.  Fig.~\ref{fig:PDF_CD_BW_Free}, which uses fixed  values of $\alpha$, shows that the amplitude of the exponential tail of the PDF is smaller for smaller values of $f$. On the other hand, Fig.~\ref{fig:PDF_CD_BW}, which shows the rescaled PDF for fixed values of $f$, demonstrates that the exponential tails are steeper for larger values of $\alpha$, as indicated by the expression for the exponents $\Lambda_n^{\rm BW}$ in Eq.~\eqref{eq:CD_diff_dom_Lambda_n_BW}.
From Figs.~\ref{fig:PDF_CD_BW_Free}~and~\ref{fig:PDF_CD_BW}, we conclude that the exponential tails of the PDFs of constant-drift inflation (in the broad-well regime) are significantly suppressed relative to those of the free-diffusion case, in the regime ${\cal N}/\varepsilon \gtrsim \mathcal{O}(1)$.

\begin{figure}[!t]
\begin{center}
\includegraphics[width=0.7\textwidth]{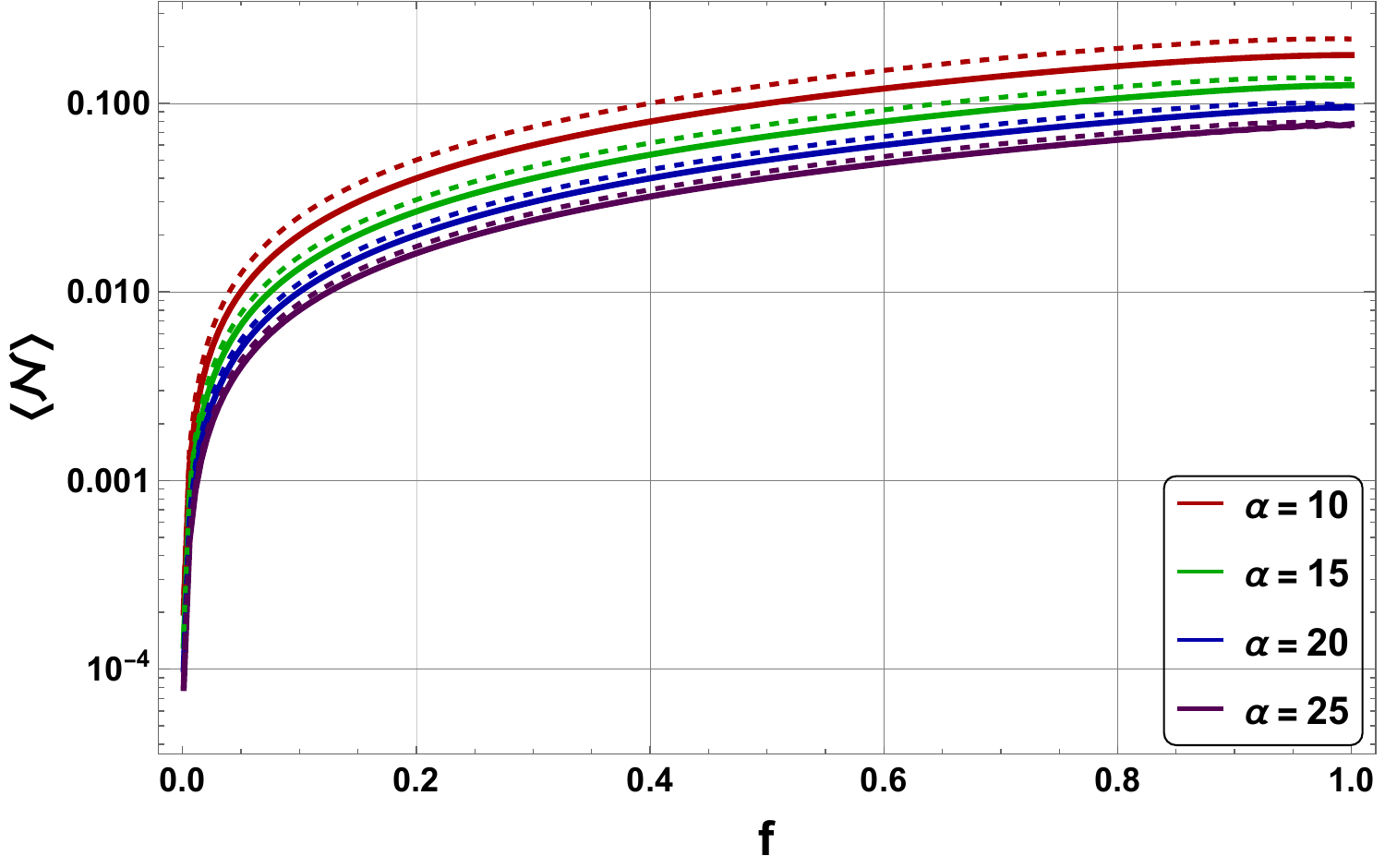}
\caption{The expectation value of the stochastic number of e-folds, $\langle {\cal N}\rangle$,  Eq.~\eqref{Nave-new} for constant-drift inflation in the broad-well regime (with $\varepsilon=1$) as a function of $f$ for $\alpha  = 10,\,15,\,20$ and $25$ (in red, green,  blue and purple, respectively). 
 The solid lines  show the numerically evaluated $\langle {\cal N}\rangle$, while the dashed lines correspond to the analytical approximation given in Eq.~\eqref{Nave-new1}.}
\label{fig:PDF_CD_BW_Navg}
\end{center}
\end{figure}

For completeness, although we do not make use of it, in order to compute  the PDF $P\l(\delta{\cal N};\,f\r)$ in Eq.~\eqref{eq:PDF_delta_N_f}, we need to first compute $\langle {\cal N} \rangle$ defined in Eq.~\eqref{eq:N_avg_f}. Following the same procedure we used to obtain ${\cal B}^{\rm BW}$ in Eq.~\eqref{eq:CD_BW_B_int2-Ed1} we obtain  
\beq
\langle {\cal N} \rangle \simeq 64 \,\varepsilon \,e^{\f{\alpha f}{2}} \sum_n \, \f{{\cal Z}_n \, \sin({\cal Z}_n\,f)}{ (\alpha^2 + 4{\cal Z}_n^2)^2} \,,
\label{Nave-new}
\eeq
which  in the large $\alpha$ regime, using 
${\cal Z}_n = (n+1) \pi \l[1-(2/\alpha)\r] $ and the summation identity~\eqref{eq:Sum_sinh}
leads to, at leading order in $1/\alpha$, 
\beq
\langle {\cal N} \rangle \simeq \f{2\,\varepsilon}{\alpha}\,f\,. 
\label{Nave-new1}
\eeq
 Fig.~\ref{fig:PDF_CD_BW_Navg} shows $\langle {\cal N}\rangle$,  numerically determined from  Eq.~\eqref{Nave-new},  as a function of $f$ for fixed values of $\alpha$,  compared to the  large $\alpha$ expression Eq.~\eqref{Nave-new1}. They agree very well, having the same shape, and differing by no more than $1\%$ for the case $\alpha=25$.

Upon comparing with Eq.~\eqref{eq:N_cl_CD}, we notice that $\langle {\cal N} \rangle \simeq N_{\rm cl}$ at leading order in $1/\alpha$. This is in stark contrast to the narrow-well limit, where $\langle {\cal N} \rangle$ was much smaller than $N_{\rm cl}$, see Fig.~\ref{fig:CD_NW_Navg_SI_Ncl}. These results have a simple physical interpretation. In the narrow-well limit, $\alpha \ll 1$, classical drift across the quantum well is subdominant compared to stochastic diffusion, \textit{i.e.}, $\l|D_\phi\r|\Delta\phi_{\rm well} \ll \Sigma_{\phi\phi}$, following Eq.~\eqref{eq:def_alpha_f}. The evolution is therefore primarily governed by stochastic dynamics, which enables the field to cross the well more rapidly, yielding a smaller average number of first-passage e-folds. In the broad-well limit, however, classical drift across $\Delta\phi_{\rm well}$ dominates over diffusion, suppressing the efficiency of stochastic transport,  thereby leading to $\langle{\cal N}\rangle\simeq N_{\rm cl}$.

Finally, to obtain the PDF $P\l(\delta{\cal N};\,f\r)$, one can simply replace ${\cal N}$ in Eq.~\eqref{PdfBWnew} with $\delta{\cal N} + \langle {\cal N} \rangle$. 

\section{Discussion and Conclusions}
\label{sec:Discussion}

We have developed an eigenvalue formulation of stochastic inflation by solving the adjoint Fokker-Planck equation using a spectral decomposition technique in association with the stochastic $\delta N$ formalism, and the   noise matrix elements computed in Refs.~\cite{Ballesteros:2020sre,Mishra:2023lhe}. In particular we are able to calculate the PDF $P({\cal N};\,\Phi)$ of the first-passage number of e-folds, ${\cal N}$, expressed as a sum over eigenmodes, with the eigenvalues, eigenfunctions, coefficients and normalization determined consistently 
with a formalism that is {\em fully self-contained} and  does not rely on complementary methods, such as the characteristic function approach used in earlier works~\cite{Pattison:2017mbe,Ezquiaga:2019ftu}.

 To demonstrate the technique we have applied our formalism to determine the PDF for two classes of features, namely, drift-free diffusion and constant-drift diffusion, showing how the former case reproduces exactly the results published in Refs.~\cite{Pattison:2017mbe,Ezquiaga:2019ftu}.  In particular we recover the 
 interesting behaviour that in addition to the well-known exponential tail for ${\cal N} \gg 1$ given in Eq.~\eqref{eq:PDF_free_tail}, there is an intermediate regime where the PDF scales as $P({\cal N}) \propto {\cal N}^{-3/2}$, Eq.~\eqref{eq:PDF_free_diff_elliptic_smallf}. This behaviour arises from the collective contribution of higher modes in the spectral sum of the PDF, Eq.~\eqref{eq:FPE_free_PDF_final}.
It modifies the shape of the PDF in between the peak and the exponential tail, and may have physical implications for the abundance of ultra-compact mini halos.

Extending the analysis to constant-drift inflation with  slow roll parameter, Eq.~\eqref{eq:eta_H}, $\eta_H = 0$, we find that obtaining the eigenvalues, which determine the form of the PDF, reduces to solving a transcendental equation, Eq.~\eqref{eq:FPE_CD_Zn_quant}, a result previously found in Ref.~\cite{Ezquiaga:2019ftu}. Although a full general analytical solution is not possible, two limiting cases are of particular physical significance: the narrow- and broad-well limits. In the narrow-well limit, since the drift, $D_\phi$, across the width of the quantum well, $\Delta\phi_{\rm well}$, is small compared to the  diffusion coefficient, $\Sigma_{\phi\phi}$, we have 
 $\alpha =  2\,\l|D_\phi\r|\,\Delta\phi_{\rm well}/\Sigma_{\phi\phi} \ll 1$ as defined in Eq.~\eqref{eq:def_alpha_f}. In this case the eigenvalues and eigenfunctions admit a controlled analytical approximation in terms of $\alpha$, leading to a consistent determination of the PDF which is simply the drift free result plus order $\alpha$ corrections, Eq.~\eqref{eq:CD_diff_dom_PDF_NW_red_final} 

In contrast, in the broad-well limit, $\alpha \gg 1$, more care is required in evaluating the PDF. This is because for a given value of $\alpha$, the solution of the eigenvalue equation, Eq.~\eqref{eq:FPE_CD_Zn_quant}, depends on the relative size of the ratio $n \pi /\alpha$, where $n$ is the mode number summed over in the PDF, Eq.~\eqref{PdfBWnew}. As $n$ runs from 0 to infinity, the form of the eigenvalues switch depending on the size of the ratio. To address this we have developed a piecewise construction of the spectrum, which goes beyond the earlier analyses~\cite{Ezquiaga:2019ftu}, by incorporating finite $\alpha$ effects, rather than relying only on the asymptotic broad-well limit $\alpha\to \infty$. This construction allows for a transition in the eigenvalues at $n \sim \mathcal{O}(\alpha/\pi)$. 
We show that the eigenvalues obtained from the piecewise construction are in excellent agreement with the numerical solutions of the transcendental equation. 

 Unfortunately,  computation of the 
PDF using the piecewise analytical construction of ${\cal Z}_n$ given in Eq.~\eqref{zncases}  introduces artificial features in the summation in Eq.~\eqref{PdfBWnew}, particularly for larger values of $f$, thereby leading to an inaccurate determination of the peak of the PDF.  
This is because the PDF is obtained from a superposition of many eigenfunctions, and is therefore highly sensitive not only to the accuracy of the individual eigenvalues, but also to the coherence of the spectrum as a function of $n$. This coherence breaks down due to the sudden switch in ${\cal Z}_n$ at $n=n_c$, and  the piecewise asymptotic construction does not preserve the approximate weighted orthogonality of the eigenfunctions, Eq.~\eqref{eq:Psi_n_othonormal_f}. Fortunately, we were able to solve exactly for the PDF numerically, and show it is well fitted by the large $\alpha$ analytic solution, Eq.~\eqref{eq:PDF_CD_BW_elliptic}. In particular, the latter provides a more accurate approximation to the numerical PDF near the peak, compared to the PDF determined using the piecewise construction. This is because it preserves the coherence of spectral structure for all values of $n$, despite providing a less accurate approximation to the individual eigenvalues for $n>n_c$.

In the narrow-well regime, we find that the exponential tail of the PDF is mildly suppressed relative to the drift-free case, while the peak remains largely unchanged. In the broad-well regime, the peak is enhanced and the tail is strongly suppressed. In the narrow-well limit the average number of first-passage e-folds is less than the classical number of e-foldings, $\langle{\cal N}\rangle < N_{\rm cl}$,  while $\langle{\cal N}\rangle \simeq N_{\rm cl}$ in the broad-well regime,  reflecting the dominance of stochastic diffusion and classical drift, respectively, across the feature.

 We note that the eigenvalue formalism developed here is quite general and can be applied to a broader class of inflationary scenarios,  such as to hilltop-type features~\cite{Boubekeur:2005zm,Kawasaki:2016pql} admitting more than one absorbing boundary, as well as to the stochastic dynamics of spectator fields~\cite{Starobinsky:1986fx,Enqvist:2008kt,Byrnes:2018txb,Chen:2024pge}. It is also directly applicable to regimes relevant for eternal inflation, where the interplay between quantum diffusion and classical drift plays a central role~\cite{Linde:1986fd,Vilenkin:1983xq,Creminelli:2008es,Rudelius:2019cfh,Tomberg:2025fku}. An important application of our framework is to constant--$\eta_H$ inflation with $\eta_H \neq 0$, which, as discussed in Sec.~\ref{sec:Inf_Dyn_cl}, frequently arises in single field models that can generate PBH forming perturbations. In this case, the spectral technique becomes considerably more involved, and a detailed analysis of the resulting eigenvalue spectrum and PDF is left for future work.

\section*{Acknowledgments}

SSM, EJC and AMG were supported by STFC Consolidated Grant Nos. [ST/T000732/1 and ST/X000672/1]. SSM is currently supported by IBS under the project code, IBS-R018-D3. AMG is supported by a Leverhulme Research Fellowship [Grant No. RF-2025-282/4], and EJC was supported by Leverhulme
Research Fellowship [Grant No. RF-2021-312]. We are very grateful to  Guillermo Ballesteros, Vincent Vennin, Christophe Ringeval and Baptiste Blachier for detailed correspondence on an earlier version of the paper.  For the purpose of open access, the authors have applied a CC BY public copyright license to any Author Accepted Manuscript version arising.

\section*{Appendices}
\appendix
\section{Formulation of the eigenvalue technique}
\label{app:ET}
\subsection{Defining the inner product for the self-adjoint Fokker-Planck operator}
\label{app:proof_self_adj_L_FP}
 Let us require that  for a suitable choice of the Sturm-Liouville weight function $w(\Phi)$, see Ref.~\cite{boas2006mathematical,SDE_Gardiner}, the adjoint Fokker-Planck operator $ {\hat {\cal L}_{\rm FP}^{\dagger}}$ becomes self-adjoint, namely
 
\beq
\langle f(\Phi), \, {\hat {\cal L}_{\rm FP}^{\dagger}} g(\Phi) \rangle_w = \langle {\hat {\cal L}_{\rm FP}^{\dagger}} f(\Phi), \,  g(\Phi) \rangle_w \, .
\label{eq:FP_op_self_adj}
\eeq
Using the expression for the inner product defined in Eq.~(\ref{eq:Inner_product}), the above expression becomes
\beq
\int_{\phi_{_A}}^{\phi_{_R}} \, {\rm d}\Phi \, f(\Phi) \l( {\hat {\cal L}_{\rm FP}^{\dagger}}g(\Phi)\r) w(\Phi) = \int_{\phi_{_A}}^{\phi_{_R}} \, {\rm d}\Phi \,  \l( {\hat {\cal L}_{\rm FP}^{\dagger}}f(\Phi)\r) g(\Phi) \,  w(\Phi) \, ,
\label{eq:FP_op_self_adj_int}
\eeq
which takes the form
$$ \int_{\phi_{_A}}^{\phi_{_R}} \, {\rm d}\Phi \, \l[ f(\Phi) \l( {\hat {\cal L}_{\rm FP}^{\dagger}}g(\Phi)\r) -  \l( {\hat {\cal L}_{\rm FP}^{\dagger}}f(\Phi)\r) g(\Phi) \r]  w(\Phi) = 0 \, .$$
Incorporating the expression for ${\hat {\cal L}_{\rm FP}^{\dagger}}$ from Eq.~(\ref{eq:Adj_FPE_Phi}), the above condition becomes
$$ \int_{\phi_{_A}}^{\phi_{_R}} \, {\rm d}\Phi \, \l[ f \l( D_\phi \, g' + \f{1}{2} \Sigma_{\phi\phi} \,  g'' \r) -  \l( D_\phi \, f' + \f{1}{2} \Sigma_{\phi\phi} \, f'' \r) g \r]  w = 0 \, ,$$
leading to 
$$  \int_{\phi_{_A}}^{\phi_{_R}} \, {\rm d}\Phi \, \f{1}{2} \Sigma_{\phi\phi}  \l( f \, g'' - f'' \, g \r) w + \int_{\phi_{_A}}^{\phi_{_R}} \, {\rm d}\Phi \, D_\phi \l( f \, g' - f' \, g \r) w = 0 \, . $$
Using the expression $fg'' - g f'' = (fg' - gf')'$ and integrating by parts, we get
$$\f{1}{2} \l[ \l( f \, g' - f' \, g \r) \Sigma_{\phi\phi}  \, w \r] \bigg\vert_{\phi_{_A}}^{\phi_{_R}} +\int_{\phi_{_A}}^{\phi_{_R}} \, {\rm d}\Phi \, \l[  \l( D_\phi -  \f{1}{2} \Sigma_{\phi\phi}' \r) w - \f{1}{2} \Sigma_{\phi\phi}  \, w' \r]  \l( f \, g' - f' \, g \r)= 0 \, .$$
The first term vanishes due to boundary conditions. For the second term to vanish the expression in the square brackets must be zero, 
which leads to a differential equation for $w(\Phi)$ 
\beq
\f{1}{w} \, \f{{\rm d}w}{{\rm d}\Phi} = \f{2 \, D_\phi - \Sigma_{\phi\phi}'}{\Sigma_{\phi\phi}}  \,.
\label{eq:diff_for_w}
\eeq
 The diffusion coefficient under the constant--$\eta_H$ approximation  is given by~\cite{Ballesteros:2020sre,Mishra:2023lhe}
\beq
\Sigma_{\phi\phi} = \f{1}{2\beta} \l(\f{H}{2\pi}\r)^2 \, , \quad {\rm with} \quad  \beta = \l\{2^{2 ( \nu - 1 )} \, \l[ \f{\Gamma(\nu)}{\Gamma(3/2)} \r]^2  \, \sigma^{2\l( -\nu + \f{3}{2}  \r)}\r\}^{-1} \,  ,
\label{eq:coeff_diff}
\eeq
where $\nu = \l| \eta_H - 3/2\r|$. In this paper, we work under the qdS approximation, $\epsilon_H \ll 1 $, for which  $H$ is nearly a constant. Hence,  the diffusion coefficient can be treated as a constant \textit{i.e.}, $\Sigma_{\phi\phi}' = 0$. Therefore, Eq.~\eqref{eq:diff_for_w} reduces to
\beq
\f{1}{w} \, \f{{\rm d}w}{{\rm d}\Phi} = \f{2 \, D_\phi }{\Sigma_{\phi\phi}}  \,  ,
\label{eq:diff_for_w_Sigma_constant}
\eeq
which has solution
\beq
w(\Phi) =  w_0\,\exp{\l(\f{2}{\Sigma_{\phi\phi}}\int_{\phi_{_A}}^\Phi  \d\Phi \, D_\phi (\Phi)\r)}   \,,
\label{eq:w_phi_final_derivation_App}
\eeq
where $w_0 = w(\phi_{_A})$ is an integration constant.  Since $D_\phi (\Phi) \equiv \pi_{\rm cl}(\Phi) = \sqrt{2} m_p \, \epsilon_H^{1/2}(\Phi)$, for constant-drift inflation we obtain
\beq
H(\Phi) = H(\phi_{_A}) \, \exp{\l[-\f{1}{2\,m_p^2} \int_{\phi_{_A}}^\Phi \d\Phi \, D_\phi(\Phi)\r]} \, ,
\label{eq:H_Dphi}
\eeq
which can also be expressed (under the constant-$\Sigma_{\phi\phi}$ approximation), using Eqs.~\eqref{eq:def_f} and \eqref{eq:def_alpha_f}, as
\beq
H(\Phi) = H(\phi_{_A}) \, \exp{\l[\l(\f{\Sigma_{\phi\phi}}{2\,m_p^2}\r) \int_0^f \d f \, \alpha(f)\r]}\, .
\label{eq:H_Dphi_f}
\eeq
In order to be consistent with the constant-$\Sigma_{\phi\phi}$ approximation (quasi-dS approximation), given Eq.~\eqref{eq:coeff_diff},  we require
\beq
 \int_0^f \d f \, \alpha(f) ~ \ll ~\f{2\,m_p^2}{\Sigma_{\phi\phi}} ~ \gtrsim ~ 10^{10}\, ,
\label{eq:Constant_Sigma}
\eeq
where the final inequality comes from the fact that $\Sigma_{\phi\phi}\approx H^2$, with  $H \lesssim 10^{-5}\,m_p$ as a consequence of the relationship between $H$ and the tensor-to-scalar ratio, $r$,~\cite{Baumann:2009ds}, and the upper bound, $r \leq 0.036$ from BICEP and Planck~\cite{BICEP:2021xfz}. 
Specialising to the case of constant-drift inflation, where $D_\phi$ is a constant, the  function in Eq.~\eqref{eq:w_phi_final_derivation_App} reduces to
$$w(\Phi) =  w_0\,\exp{\l[\f{2\, D_\phi\, \l(\Phi - \phi_{_A}\r) }{\Sigma_{\phi\phi}} \r]}   \,, $$
which can be written, using Eqs.~\eqref{eq:def_f}~and~\eqref{eq:def_alpha_f}, as
\beq
w(f) =  w_0\,\exp{\l[\l(\f{2\, D_\phi \Delta\phi_{\rm well}}{\Sigma_{\phi\phi}}\r) \, f\r]}  =  w_0 \, e^{-\alpha \, f} \,,
\label{eq:w_phi_final_CD_App}
\eeq
and Eq.~\eqref{eq:H_Dphi_f} becomes
\beq
H(f) = H(\phi_{_A}) \, \exp{\l[\l(\f{\Sigma_{\phi\phi}}{2\,m_p^2}\r)\alpha\, f\r]}\, .
\label{eq:H_Dphi_f_CD}
\eeq
To remain consistent with the constant-$\Sigma_{\phi\phi}$ approximation, given Eq.~\eqref{eq:coeff_diff}, and the fact that $f \leq 1$,  we require
\beq
\alpha ~\ll ~\f{2\,m_p^2}{\Sigma_{\phi\phi}} ~ \gtrsim ~ 10^{10}\, ,
\label{eq:Constant_Sigma_CD}
\eeq
which is trivially satisfied in the narrow-well regime  ($\alpha \ll 1$). 

\subsection{Coefficients of the eigenfunction from Fourier's trick}
\label{app:cn_Fourier_trick}

As discussed in Sec.~\ref{sec:Eigen_tech_formulation}, we need to specify the initial condition $P({\cal N}=0; \, \Phi)$, along with the absorbing and reflecting boundary conditions given in Eqs.~\eqref{eq:eigen_BCs_gen_Phi} and \eqref{eq:eigen_BCs_gen_Phi_der}, in order to completely solve the adjoint Fokker-Planck Eq.~(\ref{eq:Adj_FPE_Phi}).
The coefficients $c_n$ can be calculated conveniently using the {\em Fourier's trick}
\beq
c_n = \langle \, \Psi_n(\Phi), \, P({\cal N}=0; \, \Phi) \, \rangle_w = \int_{\phi_{_A}}^{\phi_{_R}} \, {\rm d}\Phi   \, \Psi_n(\Phi) \,  P({\cal N}=0; \, \Phi)  \, w(\Phi) \, ,
\label{eq:c_n_Fourier_trick}
\eeq
where $P({\cal N}=0; \, \Phi)$ is analogous to an initial condition for the Schr\"odinger/heat equation.
Incorporating the expression for $P({\cal N}=0; \, \Phi)$ from Eq.~(\ref{eq:ini_cond_PDF_delta_der}) into the above expression,  we obtain
\ber
c_n &=& -{\cal B} \, \int_{\phi_{_A}}^{\phi_{_R}} \, {\rm d}\Phi \,  \Psi_n(\Phi) \, w(\Phi) \, \f{{\rm d}}{{\rm d}\Phi} \, \delta_D\l(\Phi-\phi_{_A} \r) \,, \nonumber \\
&=&  -{\cal B} \int_{\phi_{_A}}^{\phi_{_R}} \, {\rm d}\Phi \, \l[  \f{{\rm d}}{{\rm d}\Phi} \Bigl( \Psi_n(\Phi) \, w(\Phi) \,  \delta_D\l(\Phi-\phi_{_A} \r) \Bigr) - \delta_D\l(\Phi-\phi_{_A} \r)  \, \f{{\rm d}}{{\rm d}\Phi}  \Bigl( \Psi_n(\Phi) \, w(\Phi)\Bigr)  \r] \,, \nonumber \\
&=&  -{\cal B} \, \Bigl[  \Psi_n(\Phi) \, w(\Phi) \,  \delta_D\l(\Phi-\phi_{_A} \r)  \Bigr]_{\phi_{_A}}^{\phi_{_R}} + {\cal B} \, \f{{\rm d}}{{\rm d}\Phi} \Bigl( \Psi_n(\Phi) \, w(\Phi)\Bigr)  \Big\vert_{\Phi=\phi_{_A}} \, ,
\eer
leading to
\beq
c_n = {\cal B} \, \f{{\rm d}}{{\rm d}\Phi}\l[ w(\Phi)\, \Psi_n(\Phi) \r] \Big\vert_{\Phi=\phi_{_A}} \, .
\label{eq:c_n_Fourie r_trick_delta_der_app}
\eeq

 Finally, we consider the determination of the parameter ${\cal B}$. Imposing  the absorbing boundary condition given in Eq.~(\ref{eq:BC_IR_PDF}) on the full PDF in the form of Eq.~(\ref{eq:PDF_gen_form_Phi})  leads to 
$$ P({\cal N}, \Phi = \phi_{_A}) \equiv \lim_{\Phi \to \phi_{_A}} \, P({\cal N}, \Phi )  = \delta_D({\cal N}) 
 \, , $$
$$\Rightarrow \lim_{\Phi \to \phi_{_A}}  \, \sum_{n=0}^{\infty} \, c_n \, \Psi_n(\Phi) \, e^{-\Lambda_n \, {\cal N}} = \delta_D({\cal N}) \, . $$
Integrating both sides wrt ${\cal N}$, we obtain
\beq
\lim_{\Phi \to \phi_{_A}} \, \sum_{n=0}^{\infty} \, \f{c_n}{\Lambda_n}  \,  \Psi_n(\Phi)  = 1  \, .
 \label{eq:c_n_sum_identity}
\eeq
Using Eq.~\eqref
{eq:c_n_Fourie r_trick_delta_der_app}, we obtain
\beq
{\cal B} =  \l[ \lim_{\Phi \to \phi_{_A}} \sum_{n} \f{\bigl(w(\Phi)\Psi_n(\Phi) \bigr)'(\phi_{_A}) \, \Psi_n(\Phi)}{\Lambda_n}\r]^{-1} \,,
\label{eq:to_B_gen}
\eeq
which is the result quoted in Eq.~\eqref{eq:To_B}.

\section{Useful formula}
\label{app:sums}
In this appendix we list the summations and integral identity  that we use in Sec.~\ref{sec:ET_Applications} and App.~\ref{app:CD_calculation}. 
\begin{eqnarray}
\sum_{n=0}^{\infty}   \, \f{\sin{ \l[ \l( 2n+1 \r)  \f{\pi}{2} \, f \r] }}{2n+1}  &=& \f{\pi}{4}\, ; \quad \text{for} \quad 0 \leq f  <
2  \, , 
\label{eq:Sum_sin_odd_1} \\
\sum_{n=0}^{\infty}   \, \f{\sin{ \l[ \l( 2n+1 \r)  \f{\pi}{2} \, f \r] }}{\l(2n+1\r)^3}  &=& \f{\pi^3}{16} \, f  \l( 1-\f{f}{2
} \r)\, ; \quad 0 \leq f  <
2  \, ,
\label{eq:Sum_sin_odd_3A} \\
\sum_{n=0}^{\infty}   \, \f{\cos{ \l[ \l( 2n+1 \r)  \f{\pi}{2} \, f \r] }}{\l(2n+1\r)^4} & = &\f{\pi^4}{192}  \l( 2 -3 \, f^2 + f^3 \r)\, ; \quad 0 \leq f  <
2  \, ,
\label{eq:Sum_cos_even_4A}
\\
\sum_{n=0}^{\infty}   \, \f{\sin{ \l[ \l( 2n+1 \r)  \f{\pi}{2} \, f \r] }}{\l(2n+1\r)^5}  &=& \f{\pi^5}{384} \, f  \l( 2 - f^2 + \f{f^3}{4
} \r)\, ; \quad 0 \leq f <
2  \, ,
\label{eq:Sum_sin_odd_5A} \\
 \sum_{m=1}^{\infty} \, \f{m \, \sin{(m\,x)}}{m^2 + b^2} &=& \f{\pi}{2} \, \f{\sinh{\l[b\,(\pi-x)\r]}}{\sinh{(\pi \, b)}}\, ; \quad 0 < x  <
2\pi  \, .
\label{eq:Sum_sinh}
\end{eqnarray}
Note that the higher order summations, Eqs.~(\ref{eq:Sum_sin_odd_3A})-(\ref{eq:Sum_sin_odd_5A}), can be obtained by integrating Eq.~(\ref{eq:Sum_sin_odd_1}), which can be found in, e.g., Ref.~\cite{Gradshteyn:2007}.

\section{Computations for constant-drift inflation}
\label{app:CD_calculation}
\subsection{Drift-dominated regime}
\label{app:CD_Drift-dom}
In Sec.~\ref{sec:ET_CD} we argued that for the case of constant drift inflation we had to work in the regime satisfying ${\cal Z}_n^2>0$, arguing that the drift-dominated regime ${\cal Z}_n^2<0$ led to inconsistent solutions which could not satisfy the boundary conditions. We show this here. In particular we consider the drift-dominated regime, ${\cal Z}_n^2<0$, which as defined in Eq.~(\ref{eq:CD_Zn_def}) corresponds to,
\ber
 {\cal Z}_n^2 = \beta_n - \f{\alpha^2}{4} < 0  \, , \nonumber \\
 \Rightarrow \beta_n < \f{\alpha^2}{4}  ~
 \Rightarrow ~ \Lambda_n < \f{1}{2} \, \f{D_\Phi^2}{\Sigma_{\phi\phi}} \, .
\label{eq:CD_Zn_driftdom}
\eer 
In this regime, since we have $\beta_n <    \alpha^2/4 $, the general solution to the eigenvalue equation Eq.~\eqref{eq:FPE_eigen_CD_f} is given by
\beq
\Psi_n(f) = e^{\f{\alpha}{2} \, f} \l[  A_n \,  e^{\l({\cal Y}_n \, f \r)} + B_n \,  e^{- \l( {\cal Y}_n \, f \r) }  \r] \, ,
\label{eq:FPE_eigen_CD_driftdom_sol_gen_a}
\eeq
or equivalently,
\beq
\Psi_n(f) = e^{\f{\alpha}{2} \, f} \l[  \tilde{A}_n \,  \cosh{\l({\cal Y}_n \, f \r)} + \tilde{B}_n \,  \sinh{ \l( {\cal Y}_n \, f \r) }  \r] \, ,
\label{eq:FPE_eigen_CD_driftdom_sol_gen_b}
\eeq
where 
\beq
{\cal Y}_n^2 \equiv - {\cal Z}_n^2 =  \f{\alpha^2}{4} - \beta_n  ~ \Rightarrow ~ {\cal Y}_n = i\,{\cal Z}_n\, .
\label{eq:CD_driftsom_Yn}
\eeq
Imposing the  absorbing boundary condition Eq.~\eqref{eq:eigen_BCs_f_Abs} at $f = 0$, namely $\Psi_n(f=0) =0$, we obtain $B_n=-A_n$, hence the eigenfunctions and their derivatives can be written as 
\ber
 \Psi_n(f) &=& A_n \, e^{\f{\alpha}{2} \, f} \,  \l[ e^{\l({\cal Y}_n \, f \r)} - e^{ -\l( {\cal Y}_n \, f \r) }  \r] = 2\,A_n \, e^{\f{\alpha}{2} \, f}  \,  \sinh{ \l( {\cal Y}_n \, f \r) }  \,, \label{eq:FPE_eigen_CD_driftdom_Abs_sol}  \\
\f{{\rm d}}{{\rm d}f} \Psi_n(f) 
 &=& A_n \, e^{\f{\alpha}{2} \, f} \l[ \l(\f{\alpha}{2} + {\cal Y}_n\r) \, e^{\l({\cal Y}_n \, f \r)} - \l(\f{\alpha}{2} - {\cal Y}_n\r) \, e^{-\l({\cal Y}_n \, f \r)} \r] \,, \nonumber \\
 &=& 2\,A_n \, e^{\f{\alpha}{2} \, f} \l[  {\cal Y}_n \,  \cosh{\l({\cal Y}_n \, f \r)} + \f{\alpha}{2} \,  \sinh{ \l( {\cal Y}_n \, f \r) }  \r]\, .
\label{eq:FPE_eigen_CD_driftdom_Abs_sol_der} 
\eer
 Similarly, imposing a reflecting boundary condition Eq.~\eqref{eq:eigen_BCs_f_Ref} at $f = 1$,
 we obtain
$$
{\cal Y}_n \, \cosh{ \l( {\cal Y}_n \r) } = -\f{\alpha}{2} \, \sinh{ \l( {\cal Y}_n  \r) \,, } 
$$
which leads to a {\em hyperbolic transcendental equation} of the form
\beq
 \tanh{ \l( {\cal Y}_n \r) } = -\f{2}{\alpha}  \, {\cal Y}_n \, .
\label{eq:FPE_CD_driftdom_Yn_quant}
\eeq
This needs to be solved  for $ {\cal Y}_n$ to determine the quantized exponents $\Lambda_n$ using Eqs~(\ref{eq:CD_Zn_def}).  No consistent solution exists for $\alpha > 0$. However, for this case we are working with $\alpha>0$ always, since if one flips the sign of $D_\Phi = \pi_{\rm cl}(\Phi) = \d\Phi/\d N$, accordingly, the sign of $\Delta\phi_{\rm well}$ also flips,  leading to $\alpha > 0$. The only way one can realise $\alpha < 0$ is by imposing $D_\phi > 0$, which means the inflaton is classically moving towards larger values of $\phi$, yet the location of the two boundaries obey $\Delta\phi_{\rm well} \equiv \phi_{_R} - \phi_{_A} > 0$. This belongs to the case of {\em hill-climbing/uphill inflation}, which is not the focus of this work.

\subsection{Derivations of expressions for quantities that appear in the PDF}
\label{app:CD_calculation_NW}
The derivation of $A_n$, Eq.~\eqref{An1} for
the case of constant-drift inflation, proceeds by first imposing the orthonormality condition given in Eq.~(\ref{eq:Psi_n_othonormal_f}),  along with the weight function in Eq.~\eqref{eq:w_f_CD}, on the eigenfunctions $\Psi_n(f)$,  given in Eq.~\eqref{eq:FPE_eigen_CD_Abs_sol}. It results in
 $$ A_n^2  \int_0^1 \, {\rm d}f  \,\sin^2{\l({\cal Z}_n \, f \r)} = \f{1}{w_0 \, \Delta\phi_{\rm well}} \, , $$
leading to
\beq
A_n^2 \,  = \l(\f{2}{w_0 \, \Delta\phi_{\rm well}}\r) \l[ 1 - \f{\sin{(2\,{\cal Z}_n)}}{2\,{\cal Z}_n} \r]^{-1} \, , 
\label{eq:CD_NW_An_1}
\eeq
where ${\cal Z}_n$ is given by Eq.~(\ref{eq:CD_Zn_def}). 
Inserting the quantisation condition, Eq.~(\ref{eq:FPE_CD_Zn_quant}),  in Eq.~(\ref{eq:CD_NW_An_1}), and using the trigonometric relation $\sin{(2x)} = 2\tan{x}/(1+\tan^2{x})$, we obtain (after a little bit of algebra) 
\beq
A_n = \sqrt{\f{2}{w_0 \, \Delta\phi_{\rm well}}} \, \l( \f{\alpha^2 + 2\,\alpha+ 4 {\cal Z}_n^2}{\alpha^2 + 4 {\cal Z}_n^2}\r)^{-1/2}  \, ,
\label{eq:CD_NW_An_2}
\eeq  
which is the same as Eq.~\eqref{An1}. The result for the narrow-well limit ($\alpha \ll 1$)  follows quickly. Recalling Eq.~\eqref{eq:epsilon_Z_NW}, it corresponds to the case 
$Z_n = [(2n+1)\pi/2] + \alpha/[(2n+1)\pi]$
then expanding Eq.~\eqref{An1} to $\mathcal{O}(\alpha)$  yields the result 
\beq
  A_n^{\rm NW}  =   \sqrt{\f{2}{w_0 \, \Delta\phi_{\rm well}}} \,  \l[ 1 - \f{\alpha}{(2n+1)^2\pi^2} \r]\, ,
\label{eq:CD_NW_An_final_app}
\eeq
which is Eq.~\eqref{eq:CD_NW_An} in Sec.~\ref{sec:ET_CD_NW}.

Turning to the evaluation of ${\cal B}^{\rm NW}$ to $\mathcal{O}(\alpha)$, we follow the same principle as above. Starting with the exact expression for ${\cal B}$, Eq.~\eqref{eq:To_B_f1}, we actually calculate ${\cal B}^{-1}$. Using Eq.~\eqref{eq:CD_NW_An} for $ A_n^{\rm NW}$ and the expression for ${\cal Z}_n$, Eq.~\eqref{eq:epsilon_Z_NW}, we expand to $\mathcal{O}(\alpha)$, ending up with a series of sine and cosine sums. These are given in    Eqs.~\eqref{eq:Sum_sin_odd_1}~and~\eqref{eq:Sum_sin_odd_3A} and, noting that the term containing the cosine summation vanishes at $f=0$, 
 we obtain  
\beq
\l(\f{w_0}{\Delta\phi_{\rm well}}\,{\cal B}^{\rm NW}\r)^{-1} \simeq \l( \f{2 \, \varepsilon}{w_0\Delta\phi_{\rm well}}  \r) \,  \l( 1 + {\cal O}(\alpha^{2}) \r)  \, ,
\label{eq:CD_NW_B_int1a}
\eeq
with the inverse of this expression, to linear order in $\alpha$, giving Eq.~\eqref{eq:CD_NW_B_int2}. 

Next, we compute  $\langle {\cal N} \rangle$ in the narrow-well regime. The underlying exact equation is given by Eq.~\eqref{eq:N_avg_f}. We obtain it as we have done above. Using Eq.~\eqref{eq:CD_diff_dom_NW_Lambda_n} for $\Lambda_n^{\rm NW}$,  Eq.~\eqref{eq:CD_diff_dom_eigen_fun_NW} for $\Psi_n^{\rm NW}$, Eq.~\eqref{eq:CD_NW_An} for $A_n^{\rm NW}$ and Eq.~\eqref{eq:CD_NW_B_int2} for ${\cal B}^{\rm NW}$ all substituted into Eq.~\eqref{eq:N_avg_f}, expanded to linear order in $\alpha$, followed by the completion of the sine sums as above,  and  some careful algebra, we obtain  Eq.~\eqref{eq:CD_NW_PDF_N_avg}.

\printbibliography

\end{document}